\documentclass[preprint2]{aastex}

\usepackage{xcolor}

\shorttitle{Formation of Brown Dwarf Binaries}
\shortauthors{Reipurth \& Mikkola}

\begin{document}

\title{Brown Dwarf Binaries from Disintegrating Triple Systems}

\author{Bo Reipurth\altaffilmark{1}
        and 
Seppo Mikkola\altaffilmark{2}
}

\vspace{0.5cm}

\affil{1: Institute for Astronomy and NASA Astrobiology Institute\\ 
          University of Hawaii, 
          640 N. Aohoku Place, Hilo, HI 96720, USA}
  \email{reipurth@ifa.hawaii.edu}

  \affil{2: Tuorla Observatory, University of Turku, V\"ais\"al\"antie
    20, Piikki\"o, Finland}
\email{Seppo.Mikkola@utu.fi}

\begin{abstract}

  Binaries in which both components are brown dwarfs (BDs) are being
  discovered at an increasing rate, and their properties may hold
  clues to their origin. We have carried out 200,000 N-body
  simulations of three identical stellar embryos with masses drawn
  from a Chabrier IMF and embedded in a molecular core.  The bodies
  are initially non-hierarchical and undergo chaotic motions within
  the cloud core, while accreting using Bondi-Hoyle accretion. The
  coupling of dynamics and accretion often leads to one or two
  dominant bodies controlling the center of the cloud core, while
  banishing the other(s) to the lower-density outskirts, leading to
  stunted growth.  Eventually each system transforms either to a bound
  hierarchical configuration or breaks apart into separate single and
  binary components. The orbital motion is followed for 100~Myr. In
  order to illustrate 200,000 end-states of such dynamical evolution
  with accretion, we introduce the 'triple diagnostic diagram', which
  plots two dimensionless numbers against each other, representing the
  binary mass ratio and the mass ratio of the third body to the total
  system mass.  Numerous freefloating BD binaries are formed in these
  simulations, and statistical properties are derived.  The separation
  distribution function is in good correspondence with observations,
  showing a steep rise at close separations, peaking around 13~AU and
  declining more gently, reaching zero at separations greater than
  200~AU. Unresolved BD triple systems may appear as wider BD
  binaries.  Mass ratios are strongly peaked towards unity, as
  observed, but this is partially due to the initial assumptions.
  Eccentricities gradually increase towards higher values, due to the
  lack of viscous interactions in the simulations, which would both
  shrink the orbits and decrease their eccentricities.  Most newborn
  triple systems are unstable and while there are 9,209 ejected BD
  binaries at 1~Myr, corresponding to about 4\% of the 200,000
  simulations, this number has grown to 15,894 at 100~Myr ($\sim$8\%).
  The total binary fraction among freefloating BDs is 0.43, but this
  assumes that all binaries are resolved and that they are all formed
  from triple systems.  However, the gradual breakup of higher-order
  multiples leads to many more singles, thus lowering the binary
  fraction. The main threat to newly born triple systems is internal
  instabilities, not external perturbations. At 1~Myr there are 1,325
  BD binaries still bound to a star, corresponding to 0.66\% of the
  simulations, but only 253 (0.13\%) are stable on timescales
  $>$100~Myr.  These simulations indicate that dynamical interactions
  in newborn triple systems of stellar embryos embedded in and
  accreting from a cloud core naturally form a population of
  freefloating BD binaries, and this mechanism may constitute a
  significant pathway for the formation of BD binaries.

\end{abstract}



\clearpage

\section{INTRODUCTION}
\label{sec:introduction}

Surveys of brown dwarfs (BDs) have demonstrated that the ratio of BDs
to stars vary between 1/6 and 1/3, possibly dependent on their
environment (e.g., Kirkpatrick et al. 2012, Scholz et al. 2013).
Three formation mechanisms for BDs have been proposed: a very low-mass
object can form if the gas reservoir is very limited (Padoan \&
Nordlund 2004, Stamatellos \& Whitworth 2009), or it can form through
dynamical ejection in small-N systems of protostellar embryos
(Reipurth \& Clarke 2001, Bate et al.  2002a, Stamatellos et al. 2007,
Basu and Vorobyov 2012), or it can form if the cloud core
photoevaporates during the collapse phase when an OB star forms nearby
(Whitworth \& Zinnecker 2004). Much debate has been exercised in favor
of one or another of these mechanisms, but as the discussions have
matured the view is emerging that all three mechanisms are likely to
operate (and in some cases even co-operate - e.g., Bate et al. 2002a),
and the question has turned to the relative productivity of the
mechanisms, which may depend on location and perhaps even cosmic time
(e.g., Whitworth et al. 2007).  Interest has shifted to the formation
and properties of BD binaries, and observations with current
techniques show that about 20\% of all BDs are resolved into binaries.
Their formation, however, is still not well understood.

In this paper, we examine the consequences of dynamical interactions
in triple systems of three identical protostellar embryos selected
from an initial mass function, and we find support for a very simple
mechanism in which triple systems of newborn protostellar embryos --
embedded in and accreting from a cloud core -- break up, ejecting a
single body while a binary recoils. The combination of chaotic
dynamics coupled with accretion from a cloud core allows various
configurations of stars and brown dwarfs, and we show that if a single
embryo falls to the center of the core and grows to become a dominant
body, then the two remaining embryos will frequently be released as a
BD binary.  We discuss the detailed processes and statistics of this
pathway to BD binary formation.

In line with the emerging view that there are multiple ways that BDs
can form, we want to clarify that we are not claiming that dynamical
interactions are responsible for all BD binaries. We demonstrate here
that this mechanism is a viable production path for BD binaries, which
in principle could be responsible for many if not all BD binaries.
Similarly, it has been argued that BD binaries may naturally form as a
consequence of turbulent fragmentation (Jumper \& Fisher 2012).  The
challenge in the coming years will be to investigate which of these,
or other, pathways to BD binary formation is the more common.

\section{THREE-BODY DYNAMICS}
\label{sec:threebody}

\subsection{Three-body Systems with Accretion}
\label{sec:accretion}

It is an interesting fact of Nature that the motion of two isolated
bodies is completely deterministic, while the addition of more bodies,
even just one, can render the motion completely chaotic. While the
gravitational two-body problem was solved already by Newton (1687), the
three-body problem has been the subject of major efforts for more than
300 years (e.g., Euler 1772, Lagrange 1778, Jacobi 1836, Hill 1878,
Poincar\'e 1892-1899). A formal solution was not discovered until the
work of Sundman (1912), which unfortunately was of no practical use in the
calculation of orbits. About a hundred years ago, the modern
era began with the first explorations of orbital integration, where
each body is moved in small steps. Burrau (1913) studied the famous
Pythagorean problem in this manner. Orbital integration lends itself
extremely well to modern computational techniques, which have
revolutionized our understanding of the three-body problem. For a
detailed discussion, see Valtonen \& Karttunen (2006) and Aarseth,
Tout, \& Mardling (2008), as well as Valtonen \& Mikkola (1991) for
astronomical applications.

\begin{figure*}
\epsscale{1.5}
\plotone{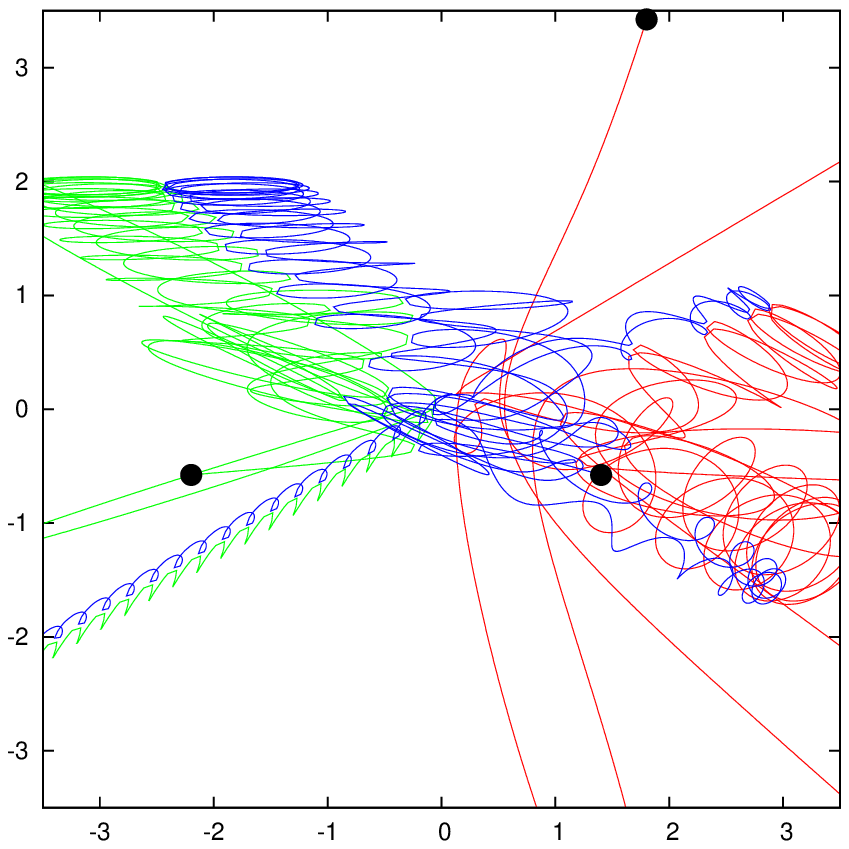}
\epsscale{1.0}
\caption{An example of the detailed motion of three bodies initially in 
a non-hierarchical configuration.  A smaller body (C, red) falls towards 
two larger bodies (A, blue and B, green), leading to a complex interplay in 
which initially A and C pair up, sending B into several large excursions, 
but after a close triple encounter A and B now form a highly eccentric 
binary which during a subsequent close triple encounter ejects C into an 
escape while the newly formed permanent binary of A and B (which is now 
more compact but still highly eccentric) recoils.
\label{orbit}}
\end{figure*}

Many subsequent studies have defined the broad characteristics of the
motion of three bodies that are initially in a non-hierarchical
configuration (e.g., Anosova 1986). There are essentially three states
of a triple system.  Initially most of the time is spent in {\em
  interplay}, during which the three bodies move chaotically with no
periodicity. Interplay is interspersed with {\em close triple
  approaches}, in which the three bodies are briefly brought close
together, and during which energy and momentum can be exchanged. The
third state is the {\em ejection}, which can only occur immediately
following a close triple approach, when one body (usually the
lightest) is ejected, while the two remaining bodies provide the
energy of ejection and as a result form a tighter bound binary system
in an orbit that is usually highly eccentric. The ejected member may
move in an approximately elliptical orbit, in which case it returns to
the binary for more interplay or another ejection.  Or it may move in
a hyperbolic orbit, in which case the ejection leads to an escape.
So {\em an ejection does not necessarily imply an
    escape}, it can be into a bound distant orbit or into an escape.
Figure~\ref{orbit} shows an example of the chaotic motions of the
interplay phase, followed by a close triple encounter and subsequent
ejection into an escape.  The motion of the three bodies in a
non-hierarchical triple system is very sensitive to even small changes
in masses.

\begin{figure*}
\epsscale{1.2}
\plotone{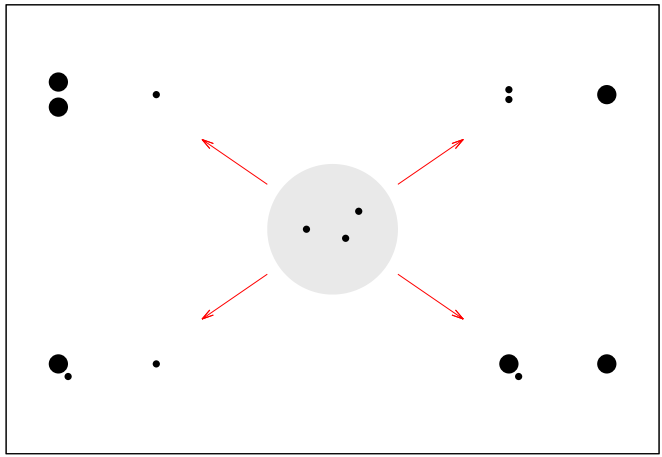}
\epsscale{1.0}
\caption{A schematic presentation of the possible pathways of dynamical 
evolution for an initially non-hierarchical triple system -- 
initially embedded in a cloud core -- of three identical stellar embryos 
which grow through competition for accretion.
\label{diagram}}
\end{figure*}

In this paper, we explore the role of a cloud core for the evolution
of a newborn triple system. In particular, we allow the three bodies
to accrete as they move around the cloud core, and this has profound
effects on the dynamical evolution of the system (e.g., Bonnell et al.
1997, Umbreit et al.  2005, Reipurth et al. 2010).
Figure~\ref{diagram} shows in broad outline the possible outcomes of
dynamical evolution of three identical stellar embryos moving inside
and accreting from a cloud core. {\em Either} one body falls to the
center of the core, rapidly growing in the process, while dynamically
keeping the two other bodies towards the outskirts of the core. If a
bound triple system results from this, it takes one of the two forms
shown to the upper right or lower left of the figure. {\em Or} two
bodies control the center of the cloud core, keeping the third body at
bay, and a resulting bound triple system will then take one of the two
forms to the upper left or the lower right of the diagram. For a
review of multiplicity among newborn stars, see Reipurth et al.
(2014).

Finally we should emphasize that we are here studying {\em isolated}
triple systems, which are thus unaffected by the presence of
additional bodies. The numerical results presented here focus on {\em
  internal} instabilities in young triple systems, and are therefore
best compared to observations of BD binary populations in T~Tauri
associations and young moving groups rather than in clusters. The same
processes are expected to take place in young clusters, but there {\em
  external} perturbations can play an additional significant role, as
discussed in Section~\ref{sec:fieldclusters}.

\subsection{Code and  Assumptions}
\label{sec:code}

\begin{figure*}
\epsscale{1.0}
\plotone{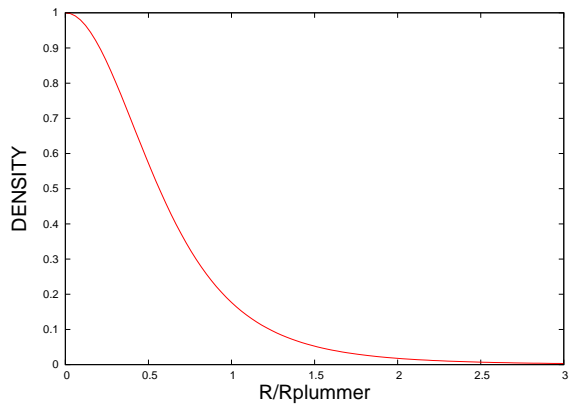}
\epsscale{1.0}
\caption{The Plummer mass distribution adopted for a cloud core. 
}
  \label{plummer}
\end{figure*}

\begin{figure*}
\epsscale{1.2}
\plotone{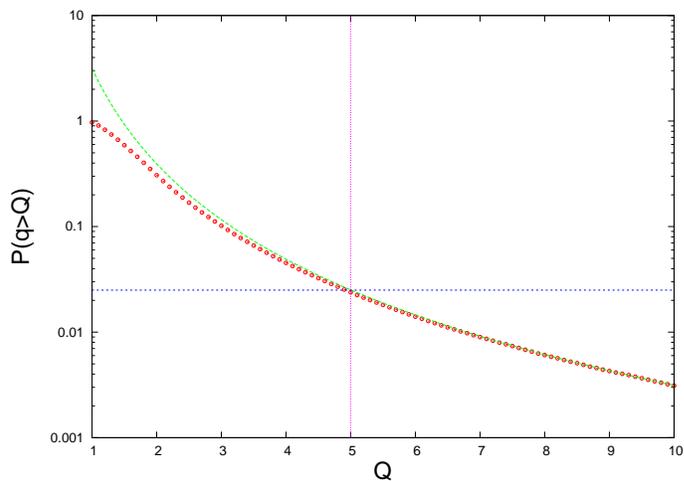}
\epsscale{1.0}
\caption{The fraction of systems in our simulations that have $q>Q$
(red dotted line) is asymptotically approaching $\pi$/$Q^3$ (thin
green line) for large $Q$, where $q$ is the ratio of separations of
the outer pair and the inner pair of a triple system, and $Q$ is a
free parameter.  The figure shows that protostars formed at random
locations inside a cloud core will in less than
2.5\% of the cases (horizontal line) have $Q>$5 (vertical line),
that is, the large majority are initially
non-hierarchical. The few systems that have hierarchical
initial configurations are eliminated.
\label{hierarchy}}
\end{figure*}

We have used a code specifically fine-tuned to
deal with the problem of three bodies moving inside a cloud core.  A
total of 200,000 simulations were performed, each for a timespan of
100~Myr.  In order to accurately calculate the frequent close
encounters, the motion of the three bodies is integrated with the
chain regularization method of Mikkola \& Aarseth (1993) that allows a
precise treatment of the gravitational force. The three bodies are
placed randomly inside a cloud core with the structure of a Plummer
sphere with radius R having a potential $\phi$(r)$=-M/\sqrt{r^2+R^2}$,
where $R$ is the radius of the Plummer core (Figure~\ref{plummer}). If
for a set of three bodies we define $q$ as the ratio of separations of
the outer pair (calculated from the most distant body to the mid-point
of the remaining two bodies) to the separation of the two closest
bodies, then one can calculate the fraction of cases that have $q$
larger than a given number $Q$. This is illustrated in
Figure~\ref{hierarchy}, which shows that the number $P(q>Q)$
asymptotically approaches $\pi$/$Q^3$.  In order to avoid introducing
hierarchical systems as initial conditions for the simulations, the
ratios of separations were not allowed to be larger than a $Q$-value of
5, and from the figure we see that this means less than 2.5\% of the
randomly chosen initial configurations were rejected. The mean initial
separations were chosen randomly between 40 and 400~AU, values
consistent with (but still poorly constrained by) observations of
embedded young stars.  The center of mass of the three bodies was then
placed at the origin which also is the center of the gas cloud.  In a
final step, the initial three-dimensional velocity vectors were
randomly chosen for each body and re-scaled so that the virial ratio
was 0.5 at the beginning of the simulations, since every initially
bound system is in that state at some time.

\begin{figure*}
\epsscale{1.4}
\plotone{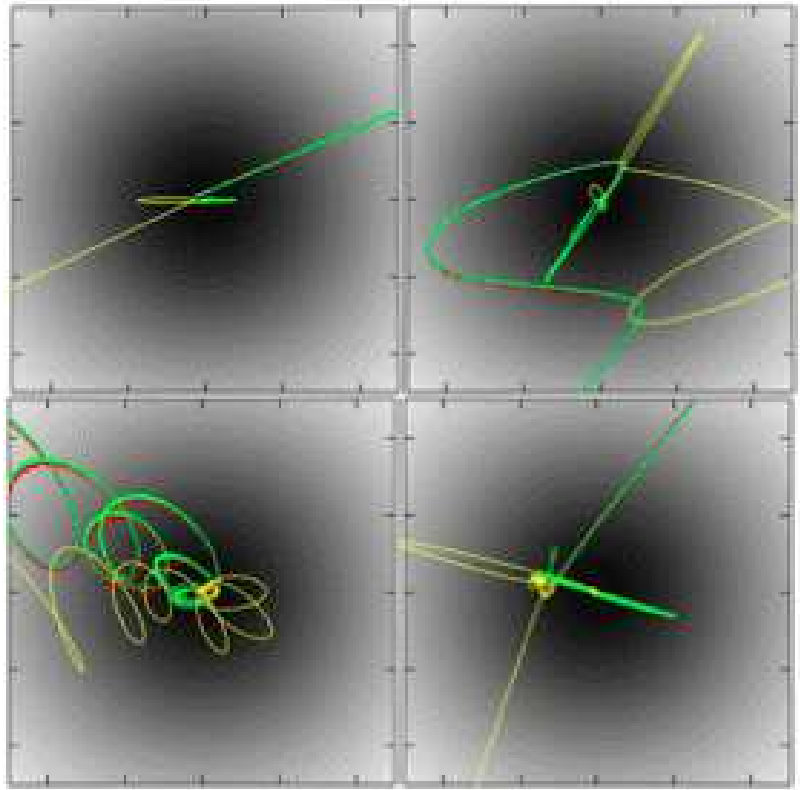}
\epsscale{1.0}
\caption{Four examples of the early large-scale dynamical behavior of a 
triple system relative to the cloud core. In three of these cases one body is 
ejected and a binary recoils. In one case (lower left) a bound triple system 
is formed, which drifts away, but it is unstable and eventually breaks up. 
The cloud core, with a Plummer density distribution and radius of 7500~AU, 
is indicated in greyscale. The width of each panel is 10,000 AU.
\label{orbitmosaic}}
\end{figure*}

The cloud cores all have a radius of 7,500~AU, a typical size
suggested by observations (Kirk et al. 2006). Core masses were
randomly chosen (i.e., probability independent of mass) in the range
from 1 to 10 M$_\odot$. The gas is assumed at rest, so accretion onto
a star causes it to slow down, because the linear momentum of the star
(mass $\times$ velocity) is kept constant.  The lack of angular
momentum of the accreting material may affect the orbits of the close
pairs, which with low angular momentum might shrink further than
determined here (Bate 2000, Umbreit et al.  2005), and with high
angular momentum gas could even expand (Bate \& Bonnell 1997, Bate
2000).  Accretion is calculated according to the Bondi-Hoyle
prescription, {\em \.{M}} $\propto \rho/(c_s^2+v^2)^{3/2}$, where $v$
is the velocity of the object relative to the gas, and the sound speed
{\em $c_s$} has a value of 0.2 km~s$^{-1}$. The cloud cores lose mass
due to accretion by the stars as well as to outflow activity, which is
assumed to cause loss of mass from the core by twice the amount that
is being accreted.  Finally, to simulate the effect of the diffuse
interstellar radiation field, the remaining gas disappears linearly
with time over a period of 440,000~yr, which is the duration of the
Class~I phase determined from Spitzer data (Evans et al. 2009).  After
the gas cloud has disappeared, the slowdown method was used to speed
up the computation (Mikkola \& Aarseth 1996).
Figure~\ref{orbitmosaic} shows four examples of the 200,000
simulations including the initial cloud core.

Evidently a full hydrodynamic calculation of the gas dynamics would be
preferable to this simplified treatment, but this would be prohibitive
when performing 200,000 simulations, as we need to obtain good
statistics of the complex and chaotic dynamical behavior of triple
systems.

We treat the stellar embryos as point sources, with no physical
extent.  This is evidently a simplification, especially at the
earliest evolutionary stages. The main effect, however, is that we can
not quantify how often stellar collisions, which certainly occur from
time to time, take place.

\begin{figure*}
\epsscale{1.2}
\plotone{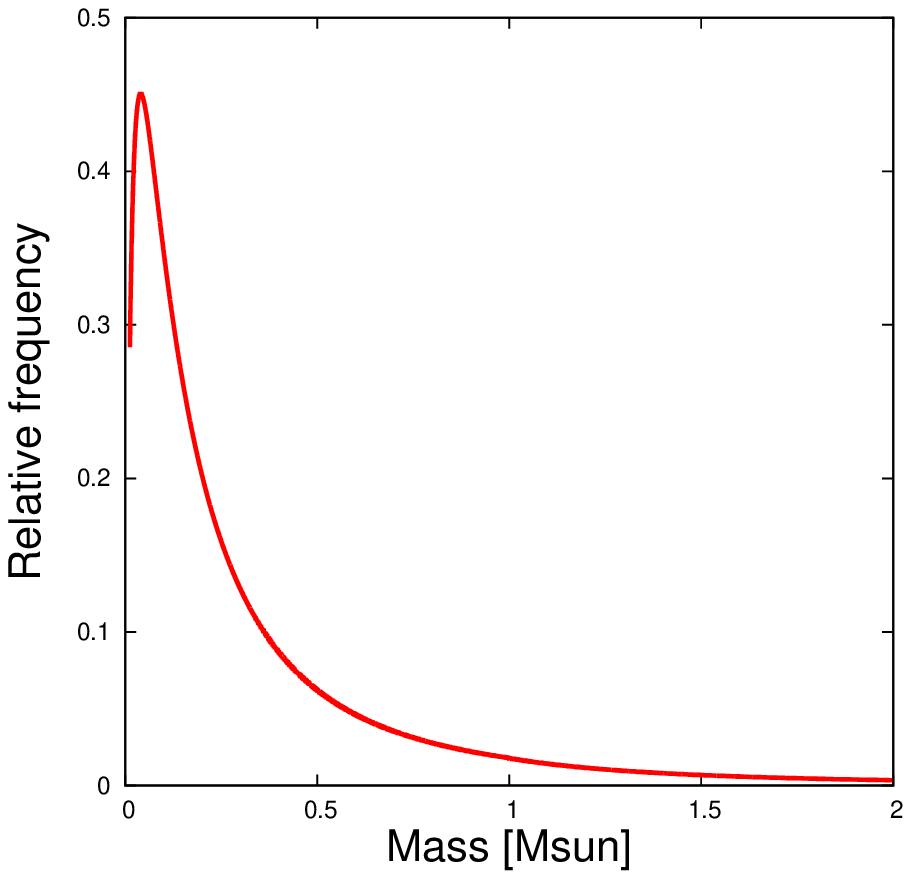}
\plotone{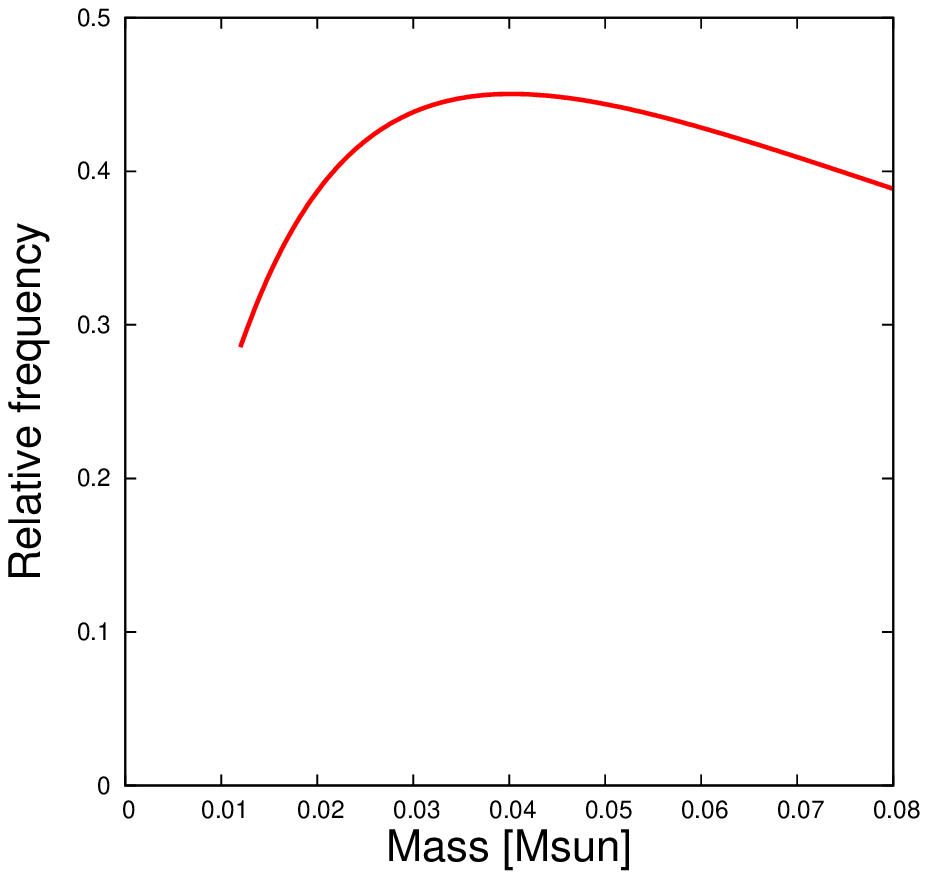}
\epsscale{1.0}
\caption{The IMF from Chabrier (2005) on a linear mass scale. We are 
selecting three identical bodies by randomly picking from this IMF 
between 0.012 and 2.0~M$_\odot$.
  \label{chabrier}}
\end{figure*}

Stellar masses are chosen from an initial mass function defined by
Chabrier (2003, 2005) which has been observationally supported by,
e.g., Alves de Oliveira et al. (2012).  At the lower end we have
truncated this IMF at 0.012 M$_\odot$, a mass that we here take to
represent the dividing line between planets and brown dwarfs.  This is
obviously quite arbitrary, but given the steep decline of the adopted
IMF at this mass range, there are very few such ``planetary mass
objects'', so our simulations are rather insensitive to this lower
limit. The maximum initial mass is set at 2~M$_\odot$, since we are
only interested in the evolution of low-mass triple systems
(Figure~\ref{chabrier}).  {\em Initially all three bodies have been
  chosen to have identical masses.}  Nothing is known about the masses
of multiple stellar embryos, so there are no empirical constraints on
whether they are identical or they differ.  However, because of the
assumption that the systems are virialized, small bodies have larger
velocities and are easily ejected to the outskirts of the cloud core
where they have very little opportunity to grow.  In other words, if
we were to choose embryo masses that were different from each other,
we would pre-determine the outcome of the simulations.  Only by
choosing three identical masses will the final masses clearly reflect
differences in their motion through the cloud core (see further
details in Section~\ref{sec:diagnostic}).  However, if Nature chooses
to form multiple embryos of different masses, the same processes will
take place, but the time-span over which they operate will be
shortened (e.g., Anosova 1986).

\subsection{Stable and Unstable Triple Systems}
\label{sec:stability}

A triple system that is non-hierarchical is inherently unstable, but
following a close triple approach and an ejection event it will either
disrupt or it will become a bound hierarchical system, which means the
system is well described as two elliptic orbits that do not cross each
other. For {\em old} stellar triple systems, the main threat to their
existence is generally considered to be external, from passing stars
that may dislodge the outermost component, most frequently due to
numerous gentle tugs rather than a single catastrophic event
(Ambartsumian 1937, Retterer \& King 1982, Weinberg et al. 1987, Jiang
\& Tremaine 2010). However, the main threat to {\em young} triple
systems is in fact internal, since the majority of newly formed triple
systems are internally unstable. The critical time for a triple system
is always the period around periastron passage of the outer component.
If at about the same time the inner system is near its apastron
passage, then the three stars are closer to each other than at any
other time. Let the inner binary components be denoted A and B, and
their orbit be numbered 1, while the outer single body is denoted C
and its orbit numbered 2.  Then the periastron distance $q_2$ from C
to the center of mass of AB is $a_2(1-e_2)$. If the $q_2$ parameter is
only a factor of a few larger than the semimajor axis $a_1$ of the
inner system, then each periastron passage of C will cause small
perturbations that after many periastron passages eventually may add
up to the point that the triple system breaks apart. This may take a
long time, and because 'a long time' can be defined in many ways,
there is no precise definition of stability.  The number of systems
deemed stable in a population may therefore vary slightly with time.
Additional effects like the Kozai resonance further complicates the
picture.

Many different formulae to predict longterm stability of a three-body
system have been proposed, all of which give slightly different results
(e.g., Huang \& Innanen 1983, Eggleton \& Kiseleva 1995, Valtonen et
al. 2008). We here use the criterion proposed by Mardling (2008).

\begin{figure*}
\epsscale{1.4}
\plotone{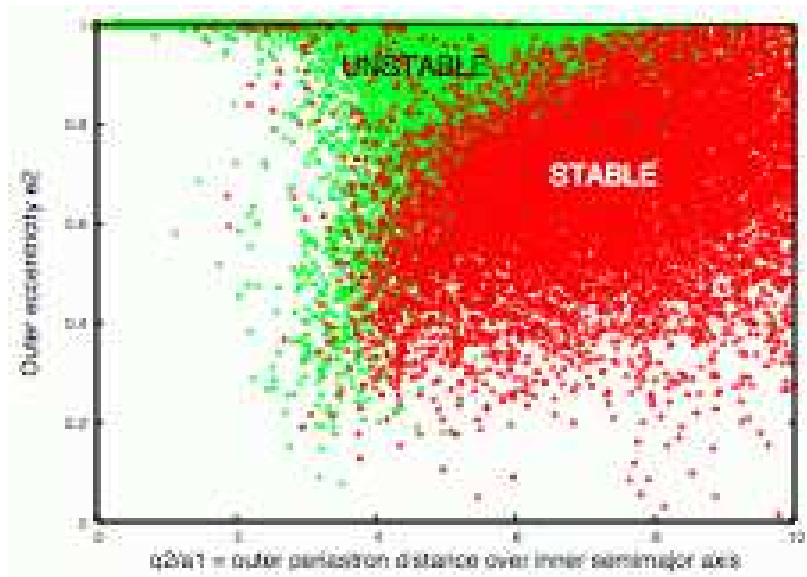}
\epsscale{1.0}
\caption{The stability criterion applied to the bound triple systems at an age
of 1 Myr. Red points (15,524) represent stable triple systems, while green 
points (62,609) represent unstable triples. Systems are stable if the distant
body does not get near enough to the inner binary to induce perturbations to
the orbital elements. See text for details. 
  \label{stability}}
\end{figure*}

In Figure~\ref{stability}, we show a plot of eccentricity $e_2$ of the
outer orbit against the $q_2$ parameter relative to the inner
semimajor axis $a_1$. The larger $q_2/a_1$ is, the less dynamical
effects will take place when the three stars are closest, and consequently
the system is more stable. But since $q_2$ = $a_2(1-e_2)$, then $q_2$
is dependent not only on $a_2$ but also on $e_2$. The larger the
eccentricity $e_2$ is, the smaller becomes $q_2$, and hence the system
becomes more vulnerable to break-up.  Physically, this is because the
binding energy of the outer orbit is small and the perturbation in
energy from the periastron passage may become of the same order of
magnitude, making disruption possible.  We show the stable and
unstable systems, classified according to the Mardling stability
criterion, as red and green crosses, respectively, and we see that the
distribution follows the behavior expected from the simple arguments
above.  A more precise statement on the stability boundary can be
found in Saito et al. (2012).

\begin{figure*}
\epsscale{1.5}
\plotone{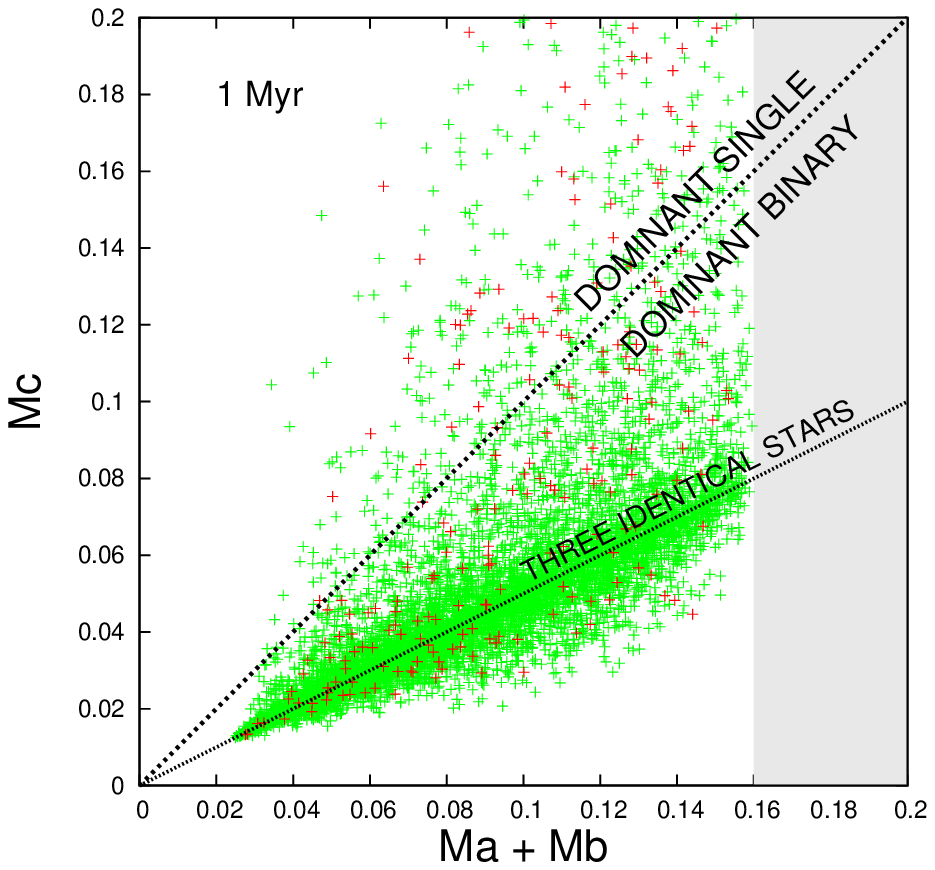}
\plotone{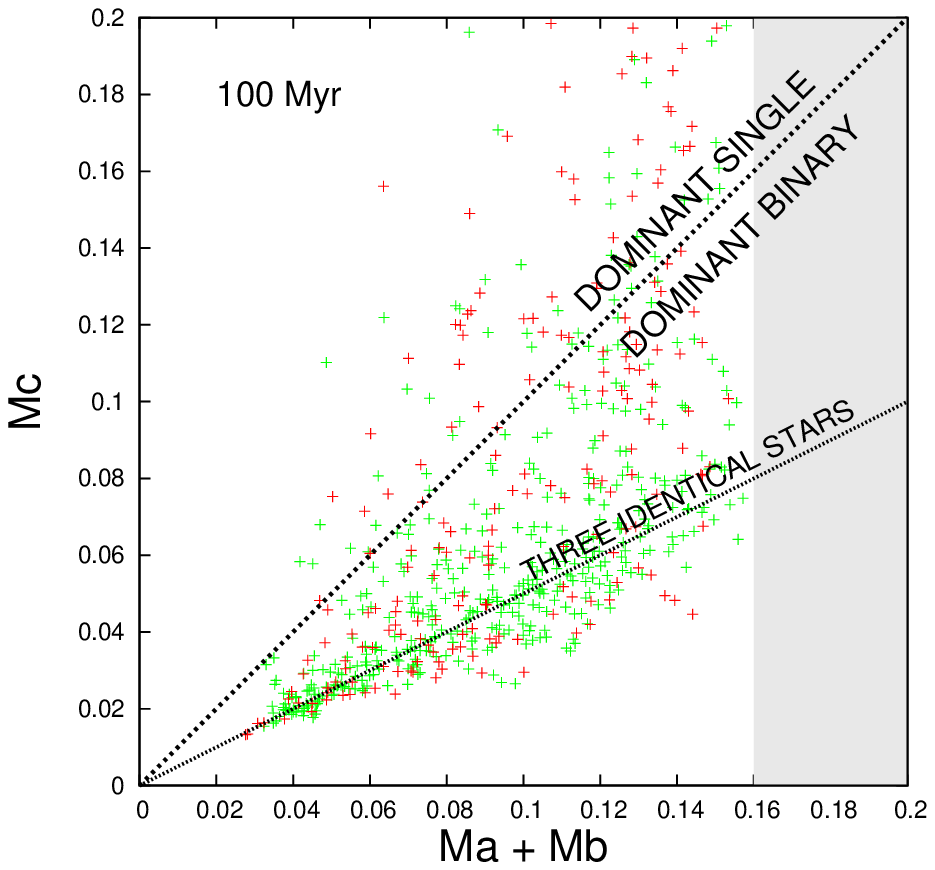}
\epsscale{1.0}
\caption{
Plot of all triple systems that remain bound at the age of 1~Myr 
and 100~Myr, respectively. Only triple systems that contain a BD binary 
are plotted. The abscissa is the binary mass, and since 
only BD binaries are shown, no objects fall in the grey region where 
binary masses would be larger than 0.16~M$_\odot$.  The ordinate is 
the mass of the third component, which can have any mass.
The locus where all three components have identical masses is indicated.
The lower right half of the diagram harbors triple systems where the binary 
mass exceeds the mass of the single component, while the upper left half
is the region where the single dominates the mass of the system.
Red points represent bound stable triple systems and green points are 
bound unstable systems. At 1~Myr many triple systems have already broken up, 
but many remain bound, although the figure shows that they are mostly unstable
(7,812 unstable vs 385 stable). At 100~Myr, most of the unstable triples 
have broken apart, releasing a BD binary into the field, and the number of 
stable (363) and unstable (472) systems are approximately equal. 
The unstable binaries that will release BD binaries are primarily located 
around the line for three identical objects, that is when a BD binary is 
created the third star that is released is also a BD or VLM object. 
\label{dominant}}
\end{figure*}

There are other more subtle influences on the stability of a triple
system. The four categories of triple systems resulting from triple
evolution with accretion and shown in Figure~\ref{diagram} are not
equally stable. In Figure~\ref{dominant}{\em a,b} we plot the mass of the
third body $M_c$ against the mass of the binary $M_a+M_b$ for all
triple systems (containing a BD binary) that remain intact at ages of
1~Myr and of 100~Myr, respectively.  A diagonal in each figure
separates systems which have a dominant single from those that have a
dominant binary. In the figure are included only binaries where both
components have masses lower than the hydrogen burning limit, hence
the cut-off at masses higher than 0.16~M$_\odot$.  The third body can
have any mass.  At 1~Myr, that is, for newly born triple systems, the
bound but unstable systems (green) far outweigh the bound stable
systems (red).  At 100~Myr, most of the unstable systems have broken
up, so the two categories of stable and unstable systems are about
equal in size.  This is very different from the state at 1~Myr, when
by far most of the unstable systems have a dominant binary. This makes
intuitive sense, because when the third body is small compared to the
binary, then the binary is much more likely to be able to perturb the
orbit of the third body when it passes through periastron.

\subsection{Dynamical Importance of a Cloud Core}
\label{sec:core}

It is a rarely appreciated fact, demonstrated by numerical
experiments, that the motion of a truly isolated non-hierarchical
triple system eventually {\em must} lead to an escape, i.e. the end
result cannot be a bound triple system. However, bound triple systems
are often observed. This is only possible if 1) the bodies are still
in an unstable configuration and have not yet disintegrated, or 2)
initial conditions accidentally created a rare stable system, or 3)
the bodies achieved a stable hierarchical configuration because they
have been formed in the presence of another body.  Except for special
initial conditions, {\em a stable triple system can only form in the
  presence of a gravitational potential, either from a cloud core or
  from additional bodies}.

\begin{figure*}
\epsscale{1.3}
\plotone{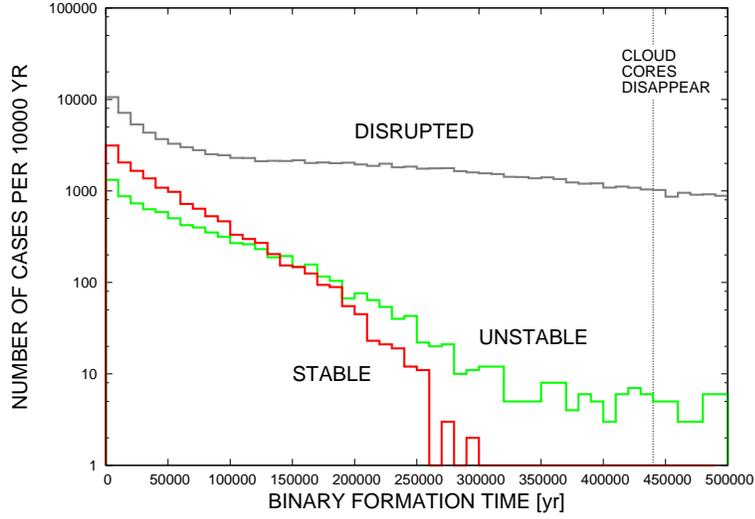}
\epsscale{1.0}
\caption{The time of binary formation. For bound systems (stable as well 
as unstable) this is the moment of
the last close triple encounter, at which point a hierarchical triple
configuration is achieved. For disrupted systems this is the moment 
of disruption. The figure shows the number of events occurring 
in time intervals of 10,000~yr, the red line indicates bound stable 
triple systems, the green line bound unstable triple systems, 
and the grey line unbound 
systems that have disrupted into a binary and a single star. 
The stars are classified according to the state 
they are in at an age of 100~Myr, when the simulations stop. The vertical line 
indicates 440,000 yr, the maximum lifetime assumed for the cloud 
cores.
\label{timeofformation}}
\end{figure*}

Figure~\ref{timeofformation} shows the time of binary formation for
the 200,000 triple systems studied here. The moment a permanent binary
forms is also the moment of the last close triple encounter (which in
some cases is also the only close triple encounter) when the triple
system transforms from a non-hierarchical configuration to a
hierarchical one, or the third body escapes. Three curves are shown,
one for bound stable triples (red), another for bound but unstable
triples (green), and the last for already disrupted systems (grey).
The classification is done for all systems at an age of 100~Myr at the
end of the simulations. The bulk of stable triples form at ages
between 10,000 and 100,000~yr, after which their formation rapidly
decreases.  After the cloud cores are gone at 440,000~yr (indicated by
a vertical dashed line), {\em no further stable triple systems form}.
The green line shows that bound triples can still form after the cloud
cores are gone, but those triples are all unstable, and will sooner or
later disrupt.

\begin{figure*}
\epsscale{1.3}
\plotone{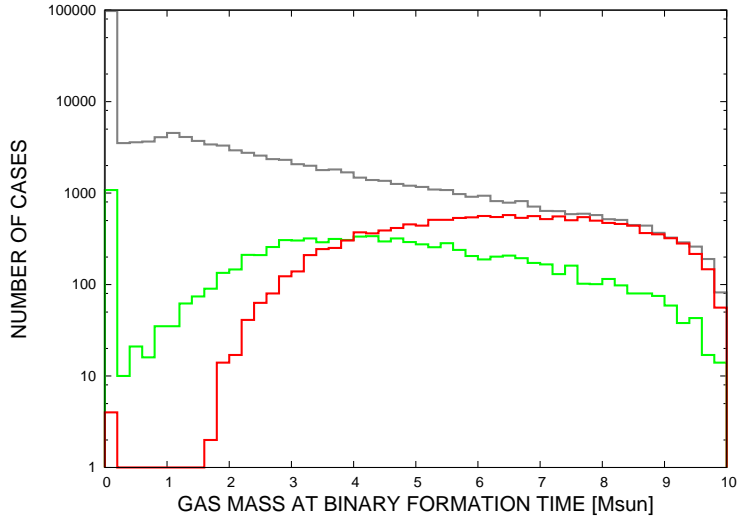}
\epsscale{1.0}
\caption{The core mass remaining at the time when the triple systems have 
their last close triple encounter and the binary becomes bound, resulting 
in either a bound stable (red) or bound unstable (green) system or a 
disrupted system (grey). As the figure illustrates, stable triple systems
are formed only in the presence of a cloud core. The few stable systems that
are indicated as having been formed with near-zero gas mass are misclassified 
by the stability criterion, they are unstable and will eventually disrupt.
  \label{coremassremaining}}
\end{figure*}

Figure~\ref{coremassremaining} shows the core mass that remains at the
time of the last close triple encounter when a binary is formed. For
each of the 200,000 simulations, an initial core mass is randomly picked in the
range from 1 to 10~M$_\odot$. So a core may have a low mass at the
time of the last triple encounter either because it was low from the
beginning, or because the core lost mass prior to that moment. The
highest mass cores preferentially form either bound stable triples
(red) or the systems break up (grey). For lower and lower masses, more
and more of the bound hierarchical triple systems that form are
unstable (green), and for core masses less than $\sim$4~M$_\odot$ the
formation of unstable triples dominate over the stable triples.

\section{THE TRIPLE DIAGNOSTIC DIAGRAM}
\label{sec:diagnostic}

\begin{figure*}
\epsscale{1.6}
\plotone{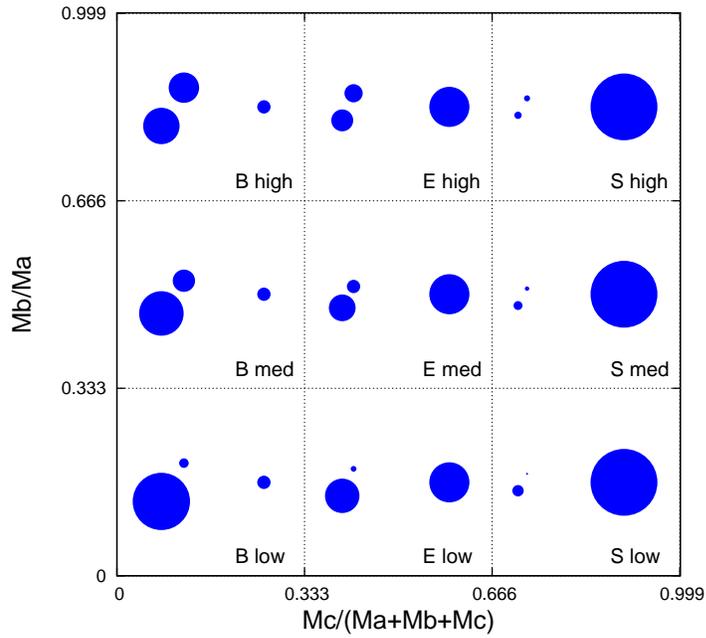}
\epsscale{1.0}
\caption{Location of different types of triple systems in the triple 
diagnostic diagram, which is useful for characterizing large populations 
of triple systems by plotting two dimensionless numbers against each other. 
The ordinate represents the mass ratio of the binary within the triple system,
while the abscissa represents the mass of the single in the triple system 
as a fraction of the total mass of the triple system. Systems to the left 
are dominated by binaries (B), and to the right by singles (S), while 
binaries and singles in the middle are roughly equal (E). Systems with high
mass ratio binaries are near the top, and with low mass ratio binaries near
the bottom.
\label{diagnostic}}
\end{figure*}

The challenge of analyzing the outcome of 200,000 simulations is
obvious.  In order to get a visual impression of the results we have
designed what we call {\em the triple diagnostic diagram}. This
diagram plots two dimensionless parameters against each other. In a
hierarchical triple system there will always be a binary and a single
star.  One important characteristic of a binary is its mass ratio,
which is always between 0 and 1. Similarly, the mass of the third body
is an important characteristic of a triple system, and if we normalize
this mass by the total mass of the triple system, then we have another
dimensionless parameter between 0 and 1.  Throughout the paper we will
denote the more massive component of the binary as A, and its
companion as B, while C is the single third body.  The triple
diagnostic diagram seen in Figure~\ref{diagnostic} has been divided
broadly into nine sections. Each resulting box shows a triple system
characteristic for that box. The three upper boxes contain all the
systems with high mass ratio binaries, while medium mass ratios are
found in the middle, and low mass ratio binaries are at the bottom. In
the three left boxes, the triple systems are dominated by the binary
systems (B), whereas the three boxes to the right are dominated by
massive singles (S), and in the middle are found systems where the
single and the binary have approximately equal masses (E). In the
following we will use terms like 'high-B systems' or 'low-S systems'
to describe certain types of triple systems.

\begin{figure*}
\epsscale{1.4}
\plotone{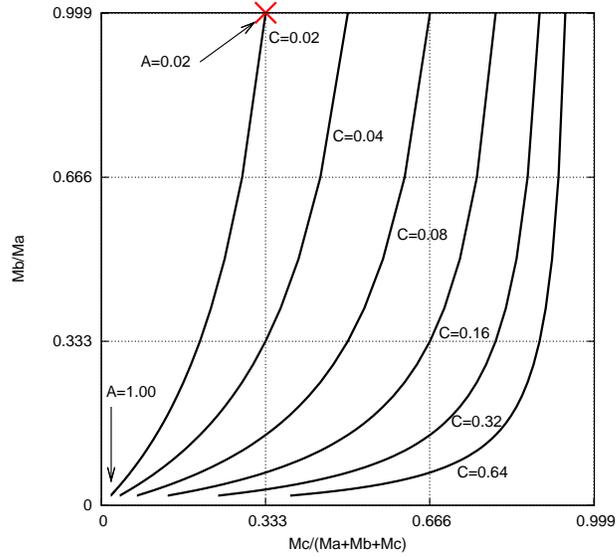}
\epsscale{1.0}
\caption{Evolution of triple systems with accretion in the triple diagnostic 
diagram. 
Note that since both axes of the diagram are ratios, then all masses can be 
scaled up or down. Those listed show ranges for M$_A$ from 0.02 to 
1 M$_\odot$, M$_B$ has a fixed value of 0.02 M$_\odot$, and six values for 
M$_C$ are shown, 
from 0.02 to 0.64 M$_\odot$. The red cross indicates the starting position 
for three identical bodies.
\label{evolution}}
\end{figure*}

\begin{figure*}
\epsscale{1.4}
\plotone{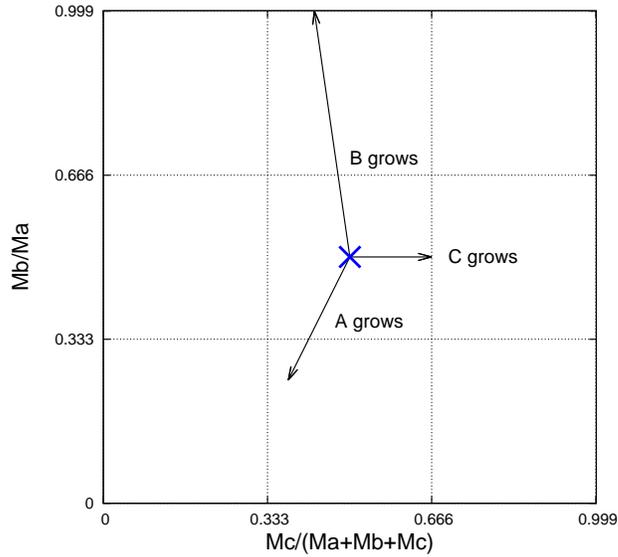}
\epsscale{1.0}
\caption{Growth of individual components in the triple diagnostic diagram.
Assume a triple system 
has evolved to reach the mass ratios A:B:C = 2:1:3 
(e.g., M$_A$=0.04, M$_B$=0.02, M$_C$=0.06 M$_\odot$), 
which corresponds to the location at the center of the diagram, 
and then assume each component is doubled in 
mass while the other two are kept constant. The three arrows show the 
new locations in the diagram. 
\label{growth}}
\end{figure*}

In Figure~\ref{evolution} the red cross indicates the location of
triple systems with all three components having identical masses.
Since the two parameters plotted are dimensionless, these triples can
be of low or high mass, as long as the masses are identical. For
purposes of illustration we shall assume all three bodies (A,B,C) to
initially have very low masses of 0.02~M$_\odot$.  If we maintain the
masses of B and C, and let A grow from 0.02 to 1~M$_\odot$ then a line
is drawn from the red cross down to the lower left corner. If we do
the same but with gradually increasing mass for the C-component, then
we get the other curves shown in Figure~\ref{evolution}.  If we
maintain the masses of A and B, and let only C grow, then we get a
horizontal line moving to the right in the diagram.  Finally, if for
fixed masses of A and C we now increase the mass of B, then the locus
in the diagram will move towards the upper left (the mass ratio will
increase so it moves up, and the total system mass increases, so it moves
left). Movements of individual components as they grow are indicated
in Figure~\ref{growth}.

\begin{figure*}
\epsscale{1.1}
\plotone{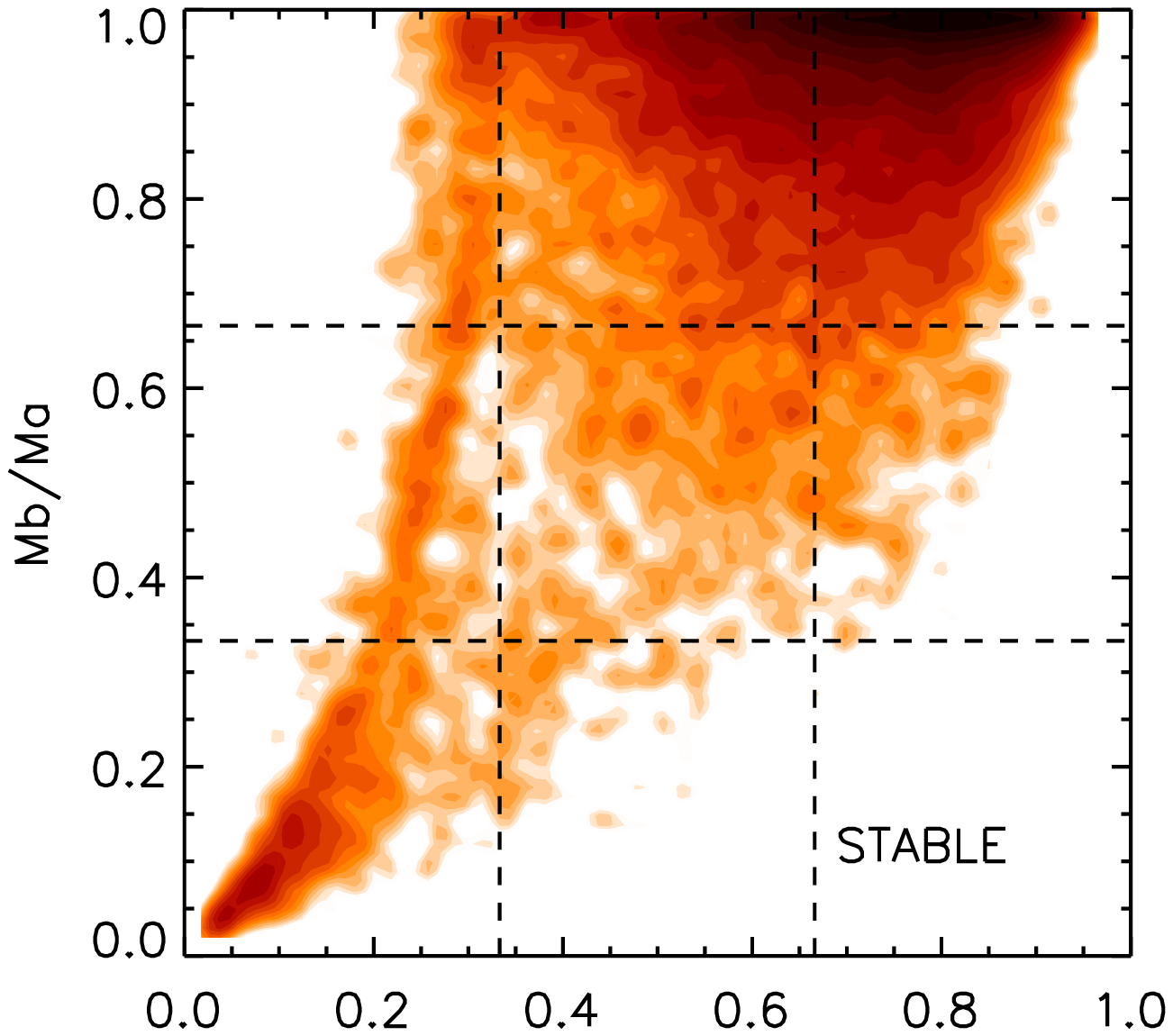}
\vspace{-1.8cm}
\plotone{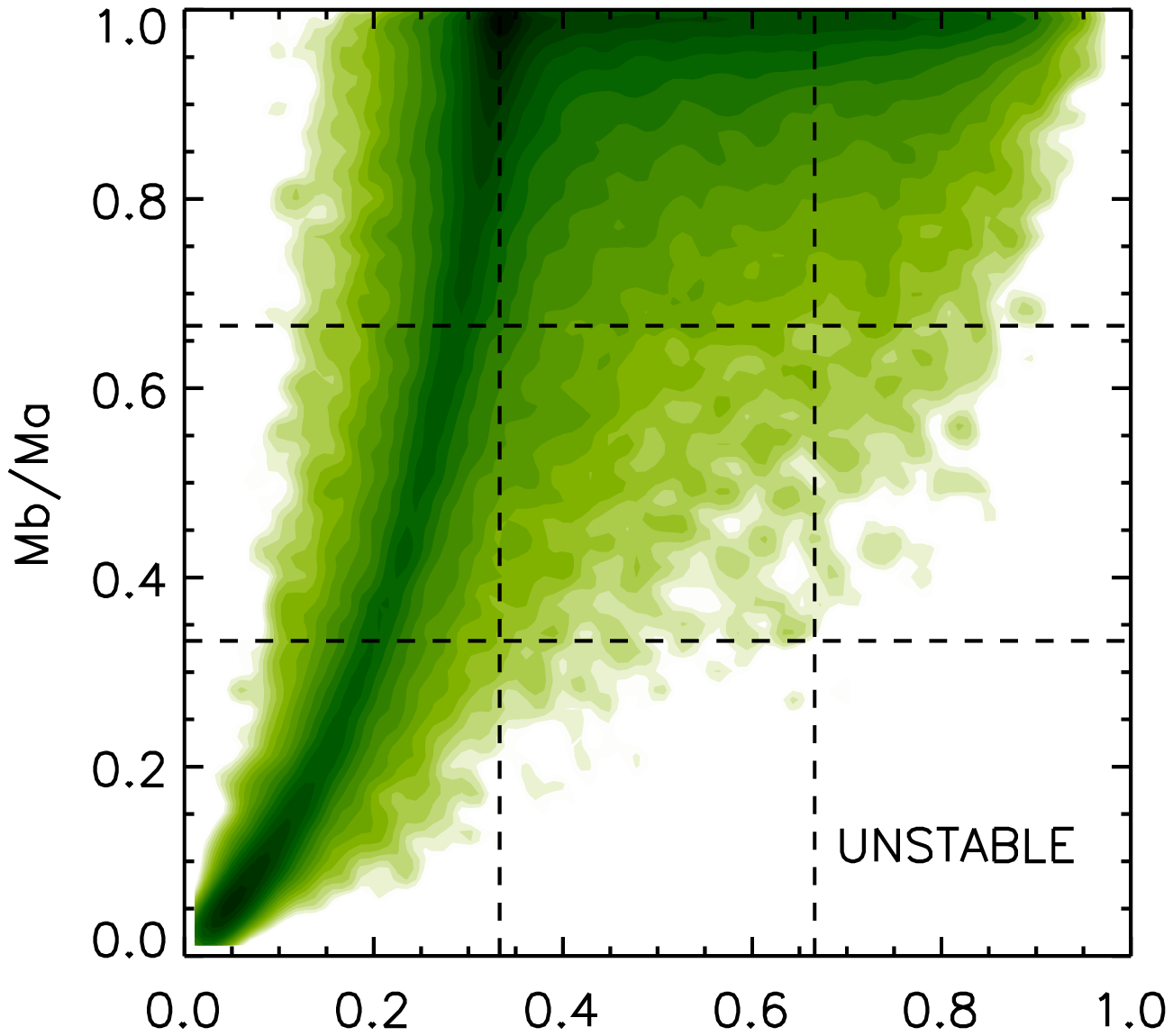}
\vspace{-1.8cm}
\plotone{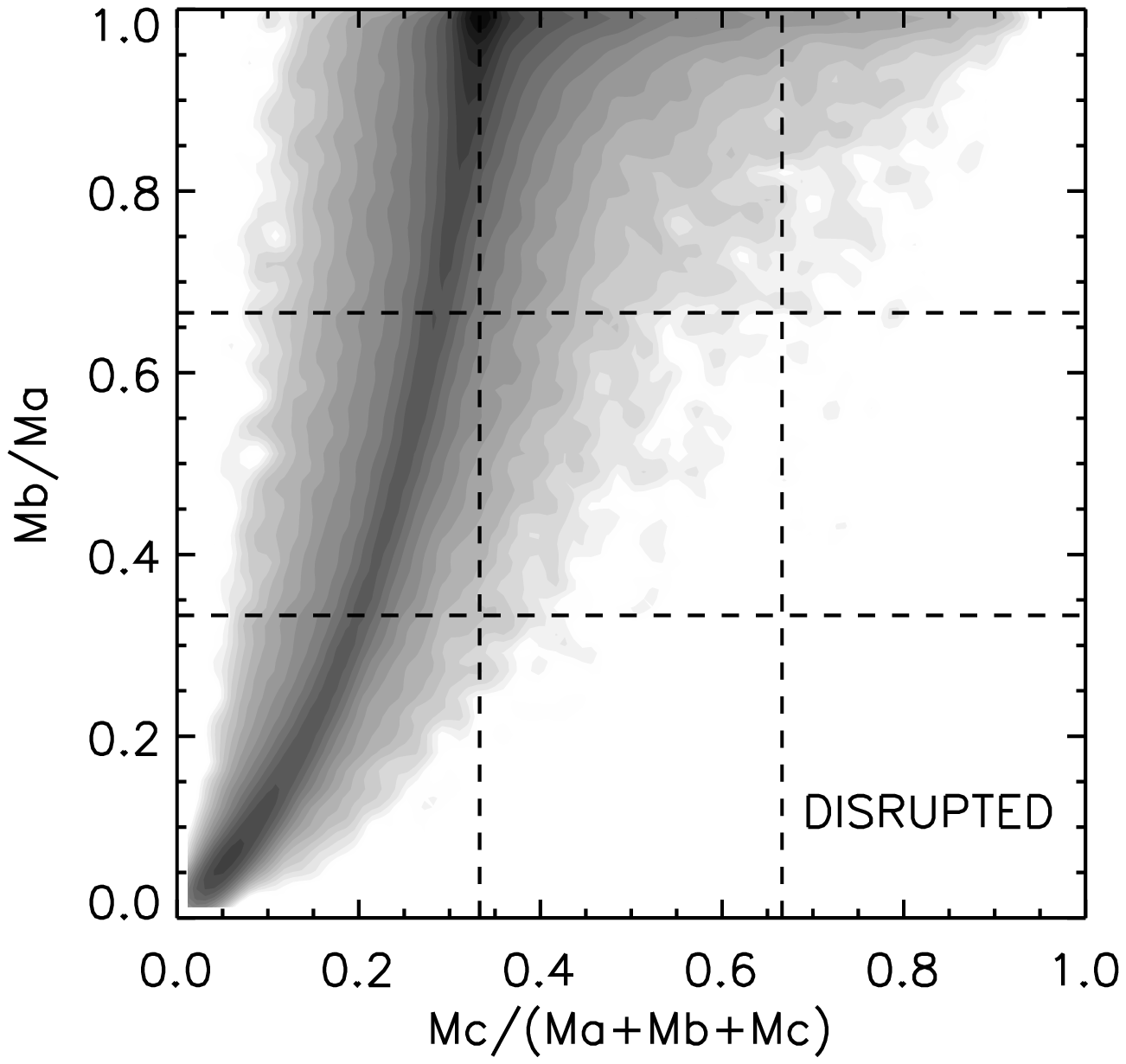}
\epsscale{1.0}
\vspace{-0.5cm}
\caption{The location of the 15,524 stable, 62,609 unstable, and 121,867 
disrupted systems in the triple diagnostic diagram at an age of 1~Myr. Since
all triple systems in these simulations were started out with three identical 
bodies, the original systems were all located at the point (0.333,1.000). Their
final location is determined by their dynamical evolution and resulting 
accretion. Evidently the resulting triple systems do not populate the triple
diagnostic diagram uniformly, but they have clear preferential locations, 
with important differences for the stable, unstable, and disrupted systems. 
The plots show smoothed surface density of points on a logarithmic intensity 
scale. Viscous interactions are not included in these simulations, and so
observations may find triple systems outside these distributions.
\label{threediagnostic}}
\end{figure*}

In Figure~\ref{threediagnostic} we show the location in the triple
diagnostic diagram of the 15,524 stable, 62,609 unstable, and 121,867
disrupted systems resulting from our 200,000 simulations at an age of
1~Myr. All these systems started out as three identical bodies (with
masses chosen from a Chabrier IMF), so they all started out at the
same point (0.333,1.000) in the triple diagnostic diagram. The chaotic
dynamical evolution of the systems lead to diverse accretion
histories, and thus to different end-locations in the diagram. The
first result we learn from looking at these plots is that there are
regions that are not populated, in other words, triple dynamics plus
accretion cannot form all types of triple systems, most notably there
is an almost complete absence of E-low and S-low systems (see
Figure~\ref{diagnostic}).  If observed triple systems fall in those
areas, then additional physical processes are required (see below and
Section~\ref{sec:limitations}).  Secondly, although similar-looking at
a quick glance, the three types of systems populate the diagram
differently.

For the stable systems, there are broadly speaking three areas of the
diagram that are particularly populated. The left edge of the
distribution forms a curved line from (0.333,1.000) down to (0,0),
i.e. from the location of three identical bodies to the location of
one massive body with two small bodies.  These are triple systems
where only one star has accreted significantly, leaving the other two
with (close to) their initial masses. To be located on this line, the
massive star and one of the small bodies form a binary, with the other
small body forming the distant third body. The second preferred area
is in the lower left corner, where the binary consists of a massive
body with a very low-mass companion and a distant, also very low-mass,
third body. Finally, the third region is the large, triangular area in
the upper right corner of the diagram, where E-high and S-high triple
systems reside. These are systems where the third body has been the
main accretor, leaving a lower-mass binary with a mass-ratio not far
from unity.

For the unstable systems, the distribution in the triple diagnostic
diagram is, in broad terms, similar to that of the stable systems,
which indicates that the mass ratios are not the main key to stability
(eccentricity and semimajor axes are, see Sect.~\ref{sec:stability}).
The main difference lies at the left edge of the distribution, which
for the unstable systems is much closer to the left edge of the
diagram. This is the area where systems with a dominant binary and a
very low-mass third body reside, and it is intuitively clear why that
area of the diagnostic diagram is not populated in the distribution of
stable systems, since periastron passages of the small third body are
likely to alter its orbit, eventually leading to instability.

Finally, for the (very many) systems that have already disintegrated
at an age of 1~Myr, the distribution mainly deviates in the upper
right corner. Here systems with dominant singles and high-mass ratio
binaries would reside, and their relative absence indicates that such
systems are particularly stable, consistent with what we saw in the
distribution of stable systems.

These results are statistically very well established within the
limits of the specific physics included in the simulations.  Currently
the observational data are much more limited and suffer from sometimes
subtle selection effects. However, when a meaningful comparison
between simulations and observations becomes possible, it is likely
that discrepancies, perhaps even significant, may appear. If so, the
most likely cause will be the absence of viscous interactions in the
simulations, see Sect.~\ref{sec:limitations}.

\section{BROWN DWARF BINARIES: \\
ANALYSIS OF SIMULATIONS}
\label{sec:analysis}

\subsection{Overview of BD Binary Formation Mechanisms}

It is widely assumed that brown dwarfs are able to form from low-mass,
turbulent-pressure confined cores (Padoan \& Nordlund 2004).
Fragmentation of such cores naturally would lead to the observed very
low mass (VLM) and BD binaries observed, and Jumper \& Fisher (2013)
show that many observed properties of such low-mass binaries are
reproduced by the turbulent core fragmentation model. It thus seems
likely that at least some BD singles and binaries are formed this way.

Whitworth \& Stamatellos (2006) suggested that if BDs form from very
low-mass cores, then close BD binaries might form by secondary
fragmentation when molecular hydrogen dissociates. This mechanism has
not been demonstrated to work in numerical simulations, and while it
produces isolated BD binaries, it does not account for the BD binaries
found in wide orbits around main-sequence stars. The latter might then
be due to capture of isolated BD binaries by stars, but Kaplan,
Stamatellos \& Whitworth (2012) show that this is highly unlikely.

Stamatellos \& Whitworth (2009) suggested that BDs could form through
fragmentation of extended very massive disks that might exist around
Sun-like stars, and that those that remain in orbit around the primary
star would be more likely to be in BD binaries than those BDs ejected
into the field. Such disks, however, remain to be observed.

In summary, it is likely that BD binaries can be formed by several
mechanisms, and it is possible that they can even work together, e.g.
a BD binary formed through disk fragmentation can later be dynamically
ejected (Bate et al.  2002a).

\subsection{Brown Dwarf Binaries as Decay Products from Triple Disintegrations}
\label{sec:decay}

Larson (1972) advocated a dynamical view of star formation in which
small groups of newborn stars interact chaotically.  McDonald \&
Clarke (1993, 1995) suggested that BD binaries do not readily form
from purely point-mass dynamical interactions in small gas-free N-body
groups -- a result confirmed by the numerical simulations of Sterzik \&
Durisen (1998) -- but that dissipative interactions must be
involved, as would happen if for example circumstellar disks would
exert drag on passing stars. 

Sterzik \& Durisen (2003) made choices of a clump mass spectrum and a
stellar mass spectrum and from a two-step process produced an IMF
compatible with the then available observations distributed across
few-body clusters, which were followed dynamically, and from which BD
binaries were produced. Accretion and gas dynamics were ignored. 

We here propose that a common pathway for the formation of brown dwarf
binaries is through the early breakup of a triple system of stellar
embryos, and we explore the statistics and time scales for the related
dynamical pathways.

\subsection{The Separation Distribution Function at 1 and 100 Myr}
\label{sec:sepdist}

\begin{figure*}
\epsscale{1.4}
\plotone{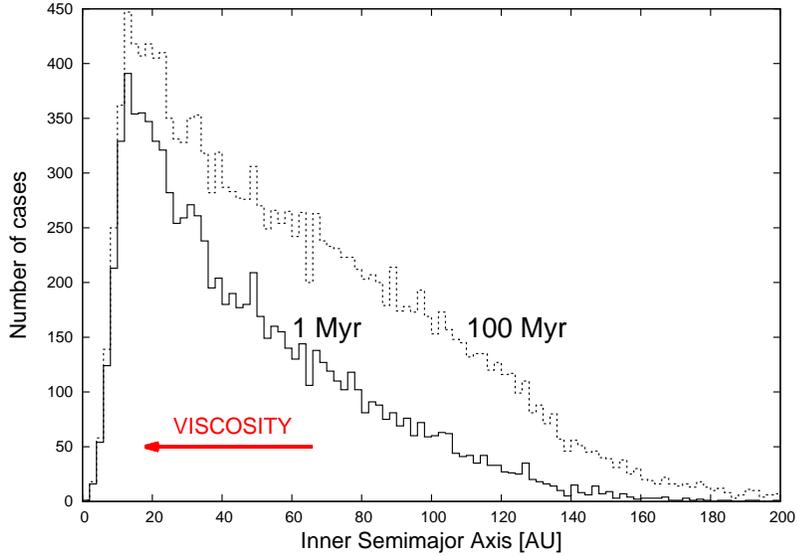}
\epsscale{1.0}
\caption{Semimajor axis distribution of ejected brown dwarf binaries 
at 1~Myr and 100~Myr. If a brown dwarf binary is observed to have a 
semimajor axis larger than $\sim$300 AU then it must be a triple system 
with an unresolved component, or a former triple system in which the close
pair merged. In that case the third body can be ejected to distances 
larger than 100,000~AU. These simulations do not include 
the effects of viscous interactions, which will tighten binaries, and bring 
the peak towards lower separations.
\label{semimajorclose}}
%
%
\end{figure*}

One of the key observational parameters for brown dwarf binaries is
the binary separation distribution function. In
Figure~\ref{semimajorclose} is shown the separation distribution of
all ejected BD-BD pairs at ages of 1 and 100~Myr.  At 1~Myr there are
9,209 ejected BD binaries, and at 100~Myr the number has grown to
15,894. At both ages, the characteristics of the distributions are the
same: a steep rise at small separations, followed by a monotonic
decrease, reaching zero at separations slightly larger than 200~AU.
At 1~Myr, the peak of the separation distribution is at 13~AU, and the
median is at 38~AU, the difference reflecting the highly asymmetric
distribution seen in the figure.  At 100~Myr, the peak of the
separation distribution is again around 13~AU, and the median is at
54~AU.  Most interestingly, very few BD binaries produced in these
simulations have a semimajor axis larger than 100~AU, and none has a
semimajor axis larger than 300~AU. The implication is that if a brown
dwarf binary is observed with a semimajor axis larger than 300~AU,
then either it formed in another way, or it must be a triple system
where one of the components is an unresolved BD binary (see, however,
Section~\ref{sec:spectroscopicbinaries}). The above results compare
well with the available observations, see Section~\ref{sec:BDprojsep}.

\begin{figure*}
\epsscale{1.4}
\plotone{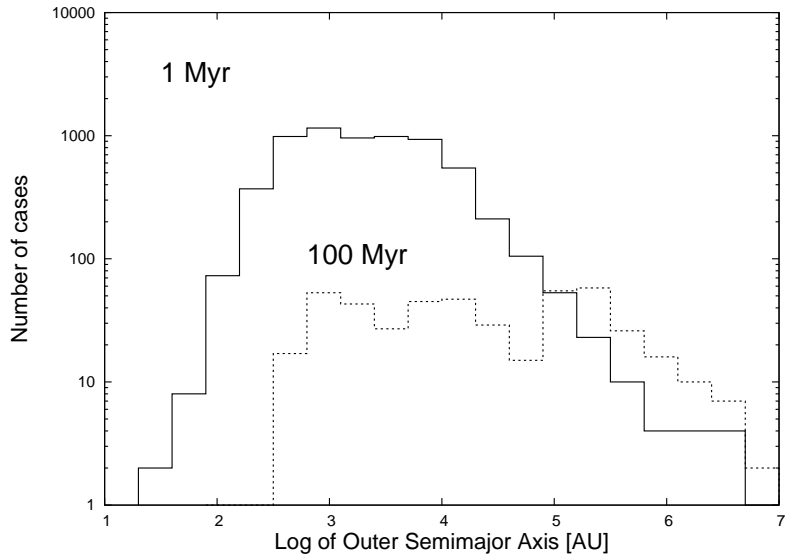}
\epsscale{1.0}
\caption{Semi-major axis distribution of the outer components in triple systems
consisting of three brown dwarfs at 1 Myr. The third body can be ejected to
very large distances and yet be kept loosely tethered to the binary.
At 100~Myr, after almost all of the unstable triples have disintegrated, 
the distribution is approximately constant in this log-diagram from 
$\sim$300~AU to several pc, which is \"Opik's Law, implying that the 
distribution is approximately $f(a) \sim 1/a$ for wide BD binaries.
\label{semimajordistant}}
\end{figure*}

At 1~Myr, there are 6,865 bound (stable and unstable) triple systems
where all components are brown dwarfs; most of these triples are
unstable (see Table~1). After 100~Myr, many of the unstable triples
have decayed, and the number of stable and unstable BD triples has
declined to just 457. The large majority of these BD triples have
outer semimajor axes much larger than 300~AU, and will at first glance
appear as wide BD binaries.  Wide and very wide brown dwarf binaries
are therefore not uncommon at young ages, but should be rare in more
evolved populations, primarily due to {\em internal} disruptions
(external disruptions from passing stars will additionally cut the
number of wide brown dwarf triples as they age).
Figure~\ref{semimajordistant} shows the separation distribution
function for the outer component of bound triple systems consisting of
three BDs on a logarithmic scale, and it is seen that the distribution
is flat from a few hundred AU to 10$^4$~AU, that is, across this large
interval it follows \"Opik's Law, see the discussion in Reipurth \&
Mikkola (2012).

\begin{figure*}
\epsscale{1}
\plotone{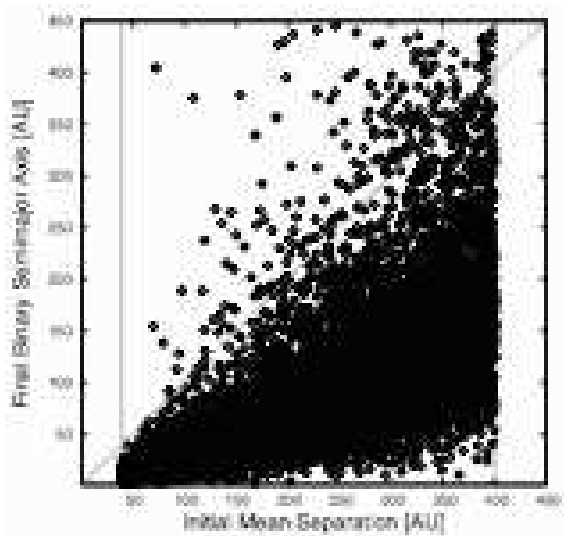}
\epsscale{1.0}
\caption{ The initial mean separations of BD triple systems (chosen to  
be between 40 and 400 AU)  have a statistical effect on the  value of 
final semimajor axes of the ejected binary systems but it is not a limit 
for it. The vast majority of BD binaries fall much beneath the dashed line
(which defines where the initial mean separation equals the final  
semi-major axis), illustrating the compression of the binary systems 
caused by the energy lost to ejection of the third body. A tiny fraction
of systems fall above the line due to low probability motions. The figure  
plots the initial mean triple separations against the final  
semimajor axes for ejected BD binaries at an age of 100~Myr. The
dashed vertical lines indicate the 40~AU lower limit and 400~AU  upper 
limit imposed on the initial mean separation of the triple  systems. 
  \label{initialseparation}}
\end{figure*}

The number of BD binaries with very close separations is significantly
smaller compared to the peak value found between 10 and 15~AU.
Dynamically, this is equivalent to the brown dwarf desert for binaries
with a solar-type star and a brown dwarf companion (Delgado-Donate et
al. 2004).  However, the statistics for these very small separations
are potentially affected by two shortcomings of our simulations.
First, our simulations assume an initial mean separation of the three
bodies in the range from 40 to 400~AU.  The absence of very close
initial mean separations could be responsible for the markedly smaller
number of close BD binaries. To search for a dependence between
initial mean separation of the newly formed non-hierarchical triple
systems with the semimajor axes of the subsequently ejected BD
binaries, we plot in Figure~\ref{initialseparation} the initial mean
separation of BD triple systems against the semimajor axes of the
ejected BD binaries.  The figure illustrates the well known fact that
binaries that formed from triple systems generally become tighter in
order to gain the energy needed to eject the third body into a distant
orbit or an escape. It also shows that there is indeed a statistical
dependence of semimajor axes on the initial mean separation, but it is
not a strong dependence, and is not likely to be responsible for the
'desert' seen at close separations.  Second, and perhaps more
importantly, our simulations do not take into account that the
numerous binaries ejected very early, during the protostellar phase,
are likely to harbor circumbinary gas, which will shrink some of their
orbits due to viscous interactions during periastron passages, thus
producing BD spectroscopic binaries.  And the binaries that remain in
the core will shrink due to gas drag (e.g., Stahler 2010).  Viscosity
thus will move the peak of the separation distribution function to
smaller values.

\subsection{Mass Ratios}
\label{sec:massratios}

An easily observed property of resolved brown dwarf binaries is the
flux ratio at various wavelengths, which if observed over a large
enough wavelength range can give the luminosity ratio of the
components. In principle, this can be converted into a mass ratio,
although in practice the degeneracy in the mass-luminosity-age
relation for brown dwarfs makes this a rather uncertain step.
Observations of VLM and BD binaries show a strong preference for high
mass ratios (e.g., Allen 2007).

\begin{figure*}
\epsscale{1}
\plotone{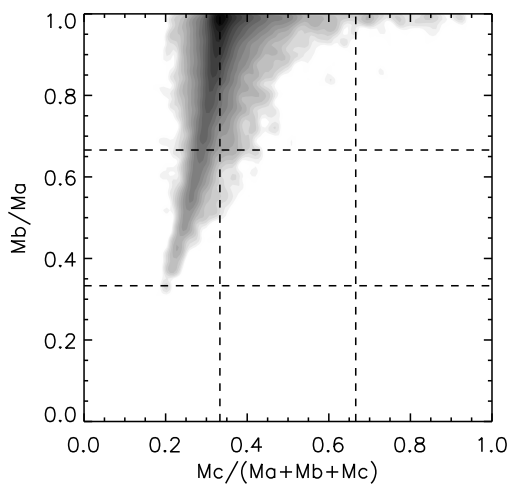}
\epsscale{1.0}
\caption{Triple diagnostic diagram for the 15,894 BD binaries that have 
resulted from the breakup of triple systems during 100 Myr. 
See text for details. 
\label{ejecteddiagnostic}}
\end{figure*}

In Figure~\ref{ejecteddiagnostic} we show the triple diagnostic
diagram for the 15,894 triple systems that have disintegrated by
100~Myr, releasing into the field a binary with two brown dwarf
components. The diagnostic diagram immediately shows two key features
of brown dwarf binaries formed from ejection: the majority have mass
ratios not too different from unity, and there are relatively few BD
binaries that have been ejected from triple systems with dominant singles.

\begin{figure*}
\epsscale{1.5}
\plotone{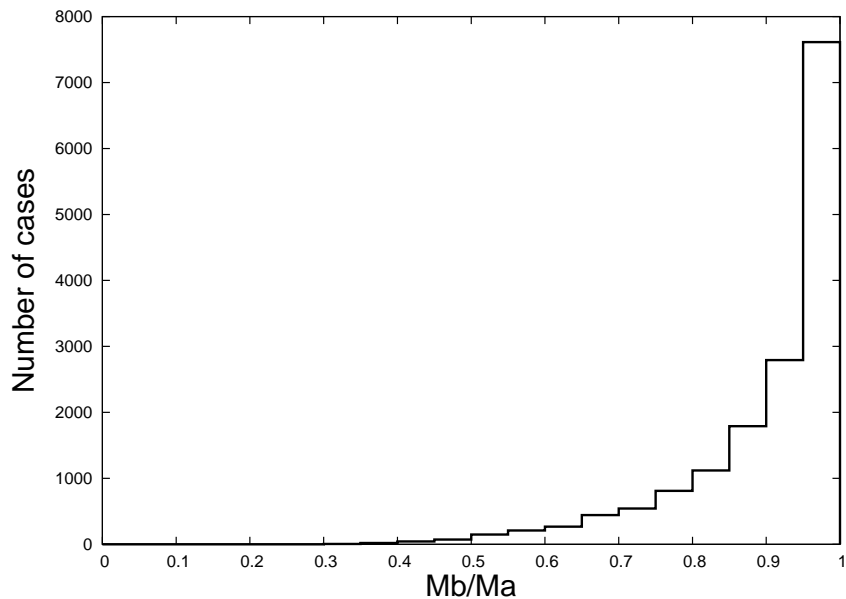}
\epsscale{1.0}
\caption{Mass ratio distribution of ejected brown dwarf binaries at 100 Myr.
The large majority of systems have mass ratios near unity, and
the mean value of the mass ratio is 0.95. 
\label{massratiodistribution}}

\end{figure*}

If we sum across the abscissa, we get the mass ratio distribution in
Figure~\ref{massratiodistribution}. The distribution is striking, showing
a preponderance of binaries with mass ratios near unity. No mass ratio
is found smaller than 0.3. The mean mass ratio for the 15,894 BD
binaries is 0.95, i.e. the components of ejected BD binaries are
frequently very alike.

This distribution reflects the fact that our stellar embryos start out
with identical masses. If two of the embryos quickly bind together,
then their motion through the core is almost identical, and they grow
by similar amounts, so their mass ratio often remains near one, even
if they grow substantially. In the cases where the binary components
have different paths through the core and only become bound into a BD
binary at a later time, the mass ratio M$_b$/M$_a$ cannot be less than
0.15 in our simulations, since the lowest mass we consider is
0.012~M$_\odot$, and the highest mass we consider as a brown dwarf is
0.08~M$_\odot$.

\begin{figure*}
\epsscale{1.5}
\plotone{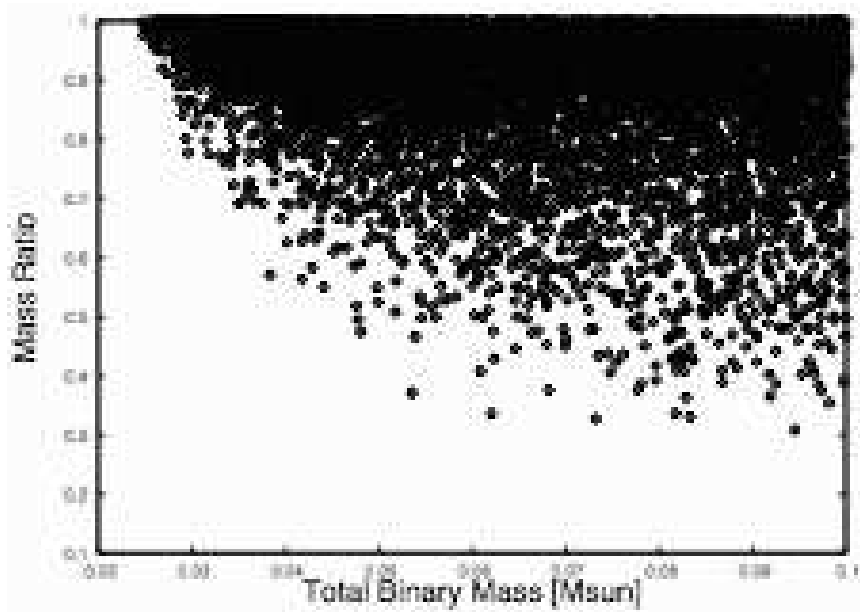}
\epsscale{1.0}
\caption{Mass ratios as function of total mass (Ma+Mb) for
ejected brown dwarf and VLM binaries. This distribution
is a natural consequence of the fact
that there are no bodies with masses less than 0.012 in the simulations: for a
binary mass of 0.024, the two smallest bodies are 0.012 and 0.012, i.e. a mass
ratio of 1. A similar, although less pronounced, effect comes from the fact 
that the IMF is rising to a peak around 0.04, after which it declines. 
\label{massratiototalmass}}
\end{figure*}

Figure~\ref{massratiototalmass} plots the mass ratio as a function of
total binary mass. The plot shows that the larger the binary mass, the
larger is the possible spread in mass ratios. There are two reasons
for this, one imposed by our assumptions, the other due to the shape
of the IMF. The lack of small mass ratio binaries at low total binary
masses is simple: if we have a primary with mass of, e.g.,
0.012~M$_\odot$, there are no bodies in our simulations with a smaller
mass, so in that case the default mass ratio is 1. A similar but
weaker and subtler effect comes from the fact that the Chabrier IMF we
are using grows rapidly to a broad peak around 0.04~M$_\odot$, so for
masses smaller than that value there is a ``deficiency'' of even
smaller masses to form small mass ratio binaries relative to masses
larger than 0.04~M$_\odot$. This is reflected in the mass ratio
distribution in Figure~\ref{massratiototalmass}
(seen by cutting vertically across the figure),
which is much steeper between 0.012 and 0.04~M$_\odot$ than between
0.04 and 0.1~M$_\odot$.

\subsection{Eccentricity}
\label{sec:eccentricity}

Another very important observational quantity is the orbital
eccentricity.  In contrast to the two previously discussed parameters,
i.e., separation and flux ratio, determination of the eccentricity
requires painstaking observations through radial velocity measurements
or high resolution imaging (e.g., Joergens et al. 2010, Dupuy \& Liu
2011), and it is consequently known much less frequently.

\begin{figure*}
\epsscale{2}
\plotone{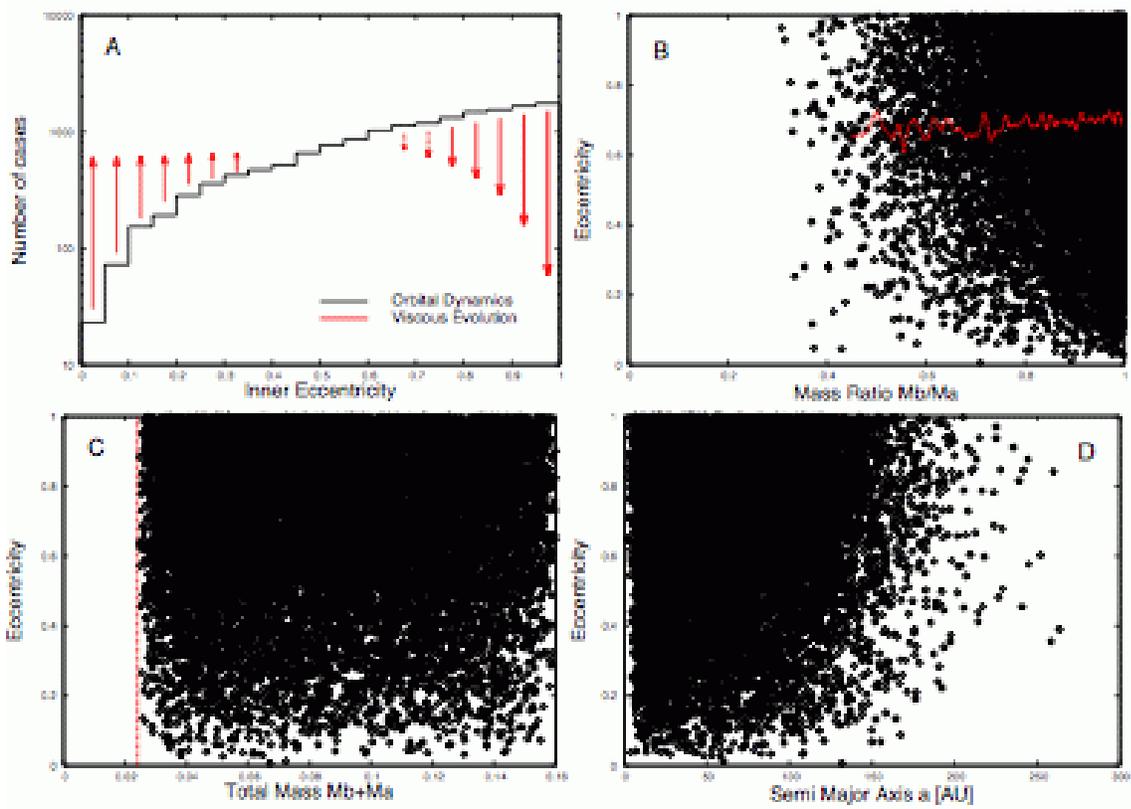}
\epsscale{1.0}
\caption{A: Eccentricity distribution of ejected brown dwarf binaries at 100 Myr. 
B: Eccentricity for ejected brown dwarf binaries at 100 Myr as a 
function of the mass ratio M$_B$/M$_A$. 
C: Eccentricity for ejected brown dwarf binaries at 100 Myr as a 
function of the total binary mass  M$_B$+M$_A$. 
D: Eccentricity for ejected brown dwarf binaries at 100 Myr as a 
function of the semimajor axis $a$. 
\label{eccentricity}}
\end{figure*}

The distribution of eccentricities for 15,894 ejected brown dwarfs at
100~Myr is shown in Figure~\ref{eccentricity}{\em a}. It forms a
monotonically increasing function towards higher eccentricities, and
the median value is 0.74. This is not consistent with the limited
information gathered so far on BD orbital eccentricities, which shows
eccentricities concentrated towards values smaller than 0.6, with only
few cases of higher eccentricity (Dupuy \& Liu 2011, Biller et al.
2013). Even taking into account observational biases, the simulations
and observations are not consistent. However, it must be recalled that
while our simulations combine orbital dynamics and mass accretion, we
do not include effects of gas dynamics. These are schematically
indicated by red arrows in Figure~\ref{eccentricity}{\em a}, and are
further discussed in Section~\ref{sec:spectroscopicbinaries}.  The
high-eccentricity binaries will decrease in number as a result of
viscous evolution, leading to an increase in low-eccentricity systems.
Consequently, the initially rising distribution due to orbital
dynamics is transformed by gas dynamics to a decreasing function.
Additionally, circularization during the pre-main sequence phase
creates circular orbits for the very shortest period binaries with
periods less than $\sim$8~days (Zahn \& Bouchet 1989).

In an attempt to understand this distribution, we examine whether the
eccentricity has any dependence on the three parameters mass ratio
M$_b$/M$_a$, total system mass M$_a$+M$_b$, and semimajor axis $a$.
Figures~\ref{eccentricity}{\em b, c}, and {\em d} plot these
parameters against eccentricity. As is evident from the figures, there
is hardly any dependence on these parameters. The only possible
exception is that BD binaries with very small mass ratios have a
somewhat weaker dependence on the eccentricity than BD binaries that
are twins or nearly identical, which shows up as a barely visible
gentle rise in a B\'ezier fit to the data points (red line in
Figure~\ref{eccentricity}{\em b}).

\subsection{Nature of the Third Bodies}
\label{sec:thirdbody}

\begin{figure*}
\epsscale{1.5}
\plotone{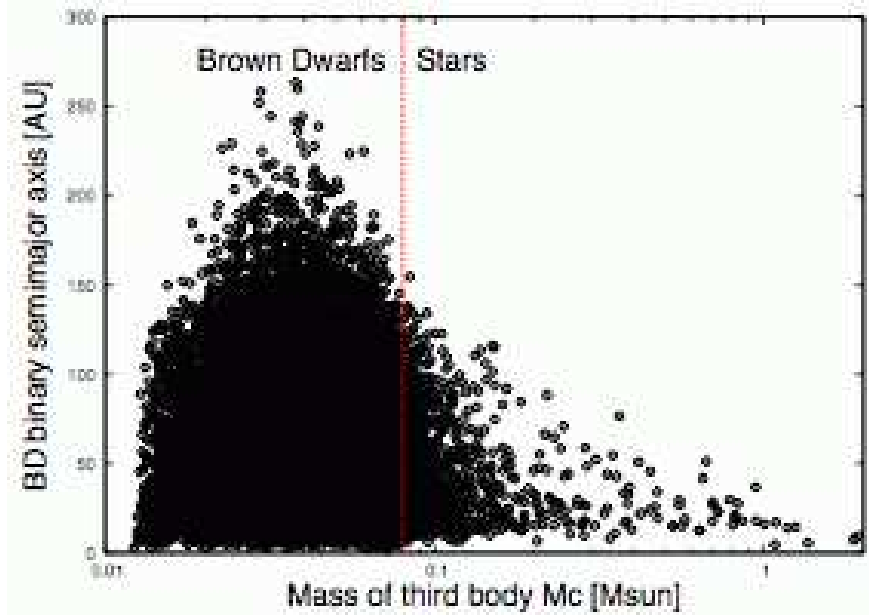}
\epsscale{1.0}
\caption{Relation between the semimajor axis of ejected brown dwarf binaries
and the mass of the third body. There are far more BD binaries that had a BD 
as the third body than a star, and they populate a much wider range of 
semimajor axes than BD binaries that were ejected from a triple system with
a star.  
\label{semimajorthirdmass}}
\end{figure*}

The brown dwarf binaries that are formed through expulsion were all
once part of triple systems, and it is of interest to ask what are the
properties of those third bodies and whether the nature of the third
body correlates with any properties of the ejected BD binaries.
Figure~\ref{semimajorthirdmass} plots the BD binary semimajor axis as
a function of the mass of the single star to which the binary was
previously bound. The main conclusion from the figure is that {\em
  ejected brown dwarf binaries are most likely to have been bound to
  another brown dwarf}. In other words, BD binaries mostly arise from
BD triple systems that disintegrated.  This partly reflects the shape
of the IMF plus the assumption that all bodies in the initial triple
systems have the same mass. The fact that most BD binaries were bound
to another BD indicates that a large number of triple systems break up
before accretion can significantly alter the masses of the components.
It also reflects the fact that in unstable triple systems, it is
predominantly the lowest-mass body that is ejected, and so if the
binary consists of two BDs, then the third body is likely to have an
even lower mass.

It is also clear from the figure that binaries ejected from stars
(as opposed to BDs) tend to have slightly smaller semimajor axes
(mean around 20-50~AU), while the mean value for BD binaries that
were bound to another BD is $\sim$50-60~AU.

\begin{figure*}
\epsscale{1.5}
\plotone{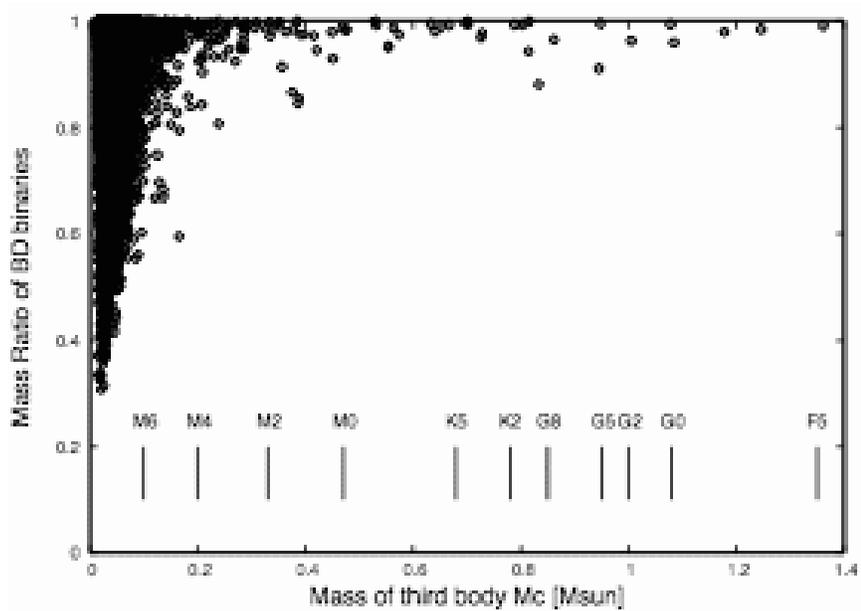}
\epsscale{1.0}
\caption{Relation between the mass ratio of ejected brown dwarf binaries
and the mass of the third body. The figure shows what is already known from
the triple diagnostic diagram, namely that S-low binaries 
(dominant single, low mass-ratio binary)
do not appear to be 
produced in triple interactions with accretion 
and an additional physical process is required to produce them.
Spectral types for main sequence stars are indicated.
\label{massratiospectraltype}}
\end{figure*}

Figure~\ref{massratiospectraltype} shows a plot of the mass ratio of
ejected BD binaries as a function of the mass of the third body.
Spectral types are indicated for different masses, based on main
sequence field stars. As already known from the triple diagnostic
diagram, triple interactions with accretion are unlikely to produce
S-low binaries, that is,  BD binaries with small mass ratios do not
originate from triple systems that include a star.

\subsection{Statistics of Binarity} 
\label{sec:statistics}

As mentioned in Section~\ref{sec:diagnostic}, the binary components
are denoted A and B, with A being the more massive, while C is the
third body, either bound or escaped. For a three-body system
consisting of stars and brown dwarfs, there are six possible
combinations: either the third component C is a star or a brown dwarf,
and for each of these two cases, A and B can be both brown dwarfs or
both stars, or A can be a star with B as a brown dwarf. These six
categories are listed in Table~1 for stable hierarchical systems (H),
unstable hierarchical systems (U), and disrupted systems (D), all at
ages of 1, 10, and 100~Myr. Brown dwarf binaries that originate from
disintegrated triple systems may either have had a brown
dwarf or a star as a third member. At 1~Myr this adds up to
4.26\%+0.35\% = 4.61\% of the initial population of 200,000 triple
systems in the simulations. At 10~Myr, this has grown to 7.40\%, and
at 100~Myr it is leveling out at 7.94\%.  As discussed in
Section~\ref{sec:thirdbody}, the majority of these brown dwarf
binaries originate from systems where the third body was also a brown
dwarf.

An important observational parameter is the fraction of brown dwarfs
that are binaries. This we can derive from Table~1, and the numbers at
an age of 1~Myr are given in Table~2. The percentage of the original
triple systems which have ejected a single brown dwarf at 1~Myr is
6.07\% and, as mentioned above, the number of ejected brown dwarf
binaries is 4.61\%. In total this means that 10.68\% of the 200,000
systems are resulting in either a single BD or a BD binary.  Of these,
the fraction of binaries are 4.61/10.68 $=$ 0.43. This is higher than
observed. However, the numbers assume that {\em every} BD
binary is resolved and identified as such. Given that many of these
binaries are very close and will be resolvable only through adaptive
optics observations or through a radial velocity study, we must expect
that a fraction of the binaries are not resolved, but appear to the
observer as single BDs. Table~2 includes the BD
binary fraction assuming that the unresolved binaries range from 25\%
to 50\% to 75\%. The actual binary fraction of 0.43 then declines to
0.32, 0.22 and 0.11, respectively.

Similar calculations at 10 and at 100~Myr yield the same binary
fractions as at 1~Myr to within the two-digit accuracy we use here.
The binary fraction is thus not age dependent.  This is because the
six categories of triple systems listed in Table~1 decay at
essentially the same rate with time.

Note that in these calculations, we are not including the BD triple
systems which remain bound. If the close BD binary is not resolved,
such triple systems will appear as two widely separated BDs (typically
$>>$200~AU).

\begin{figure*}
\epsscale{1.5}
\plotone{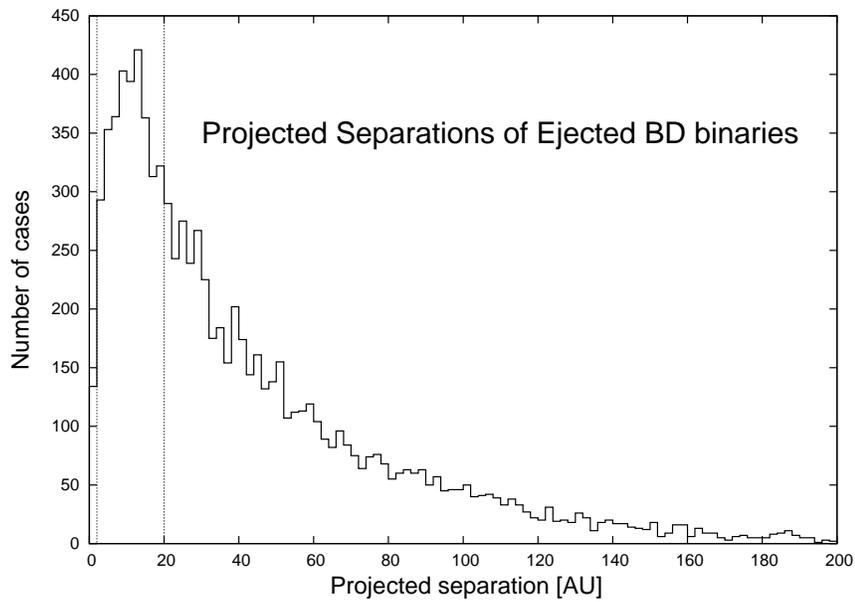}
\epsscale{1.0}
\caption{The distribution of projected separations of ejected BD binaries. 
The figure shows the distribution for 9,209 BD binaries at 1~Myr.
The distribution has a 
broad peak between 2 and 20~AU (indicated by the vertical dotted lines), 
in which interval 35\% of the ejected BD binaries can be found. 
No binary has a projected separation larger than 250~AU. Wider systems are 
common, but in that case they are (or have been) triple systems.
  \label{BDejectedprojsep6}}
\end{figure*}

At first glance, this high fraction of BD binaries does not appear
consistent with observations. In the most recent and most complete
study, Todorov et al. (2014) found a binary fraction of
0.04$^{+0.03}_{-0.01}$ among young BDs for resolved binaries with
separations $>$10~AU. With this resolution, the observations only
probe longwards of the 11~AU peak of the BD binary separation
distribution function (Figures~\ref{semimajorclose} and
\ref{BDejectedprojsep6}), and so ignores the population of
spectroscopic binaries, which are discussed in the following
(Section~\ref{sec:spectroscopicbinaries}).  Although detections of
spectroscopic BD binaries are still limited, the existing preliminary
surveys indicate that such spectroscopic BD binaries should be rather
common, Maxted \& Jeffries (2005) suggest 17\%--30\%, Basri \& Reiners
(2006) 26\%$\pm$10\%, and Allen (2007) about 20\%. The total number of
observed BD binaries is still less than the 0.43 of the simulations.
But the calculated fraction assumes all BD binaries are ejected from
triple systems.  Ongoing simulations (in prep.) of higher-order
multiples show that many more single BDs are ejected as soon as the
number of bodies in the multiple system increases, and hence the BD
binary fraction falls significantly.  Finally, since our simulations
do not include the effect of viscous interactions, a number of our
binaries will shrink into the spectroscopic binary range.  A more
detailed comparison of binary statistics between observations and
simulations must therefore await the study of higher-order multiples
and a more complete survey for spectroscopic BD binaries.

\subsection{BD Spectroscopic Binaries and Mergers}
\label{sec:spectroscopicbinaries}

Spectroscopic binaries are known to exist among the BD and VLM
population (e.g., Basri \& Martin 1999, Joergens 2008, Burgasser et
al. 2012, Clark et al. 2012).  Very few BD binaries are formed in our
simulations with semimajor axes of only a few AU, which is the realm
of the spectroscopic binaries, so evidently purely dynamical processes
do not readily form such close pairs. The additional physical process
missing in our simulations is the role of viscosity.  The motion of a
binary through the gas in a dense cloud core creates drag, and leads
to an orbital decay of the binary components (Gorti \& Bhatt 1996,
Stahler 2010, Korntreff et al.  2012).  In some cases even mergers can
take place (Rawiraswattana et al. 2012, Leigh \& Geller 2012,
Korntreff et al. 2012).  The majority of ejections and the resulting
formation of a bound binary occurs during the embedded protostellar
phase (Reipurth 2000, Reipurth et al.  2010), and at these early
stages the stars are also surrounded by significant circumstellar
disks, which interact during the periastron passages of the usually
very eccentric orbits, resulting in further orbital decay (McDonald \&
Clarke 1995, Hall et al. 1996, Bate et al.  2002b).  These processes
stop to act when first the cloud core vanishes, and later the disks
disperse, leaving the BD binary with whatever orbital parameters that
resulted from the spiral-in process.  Full SPH simulations are
required to document the role of these processes in forming the
closest BD binaries. In consequence, {\em the distribution of
  semimajor axes in Figure~\ref{semimajorclose} reflects only orbital
  dynamics, and does not include the effects of gas-induced orbital
  decay}. Similarly, the eccentricity of the viscously interacting BD
binaries will gradually diminish, as schematically indicated in
Figure~\ref{eccentricity}{\em a}.

From a practical point of view it is unfortunate that mergers are
likely to occur for some of the close binaries in triple systems, since this
implies that it is not possible to test the prediction made in
Section~\ref{sec:sepdist} that BD binaries wider than $\sim$300~AU
should be triple systems. If the close binary has suffered a merger,
then the wide BD binary will in fact only consist of two objects,
notwithstanding their origin in a triple system.

\subsection{Binding Energy as Function of Total Binary Mass}
\label{sec:bindingenergy}

\begin{figure*}
\epsscale{1.5}
\plotone{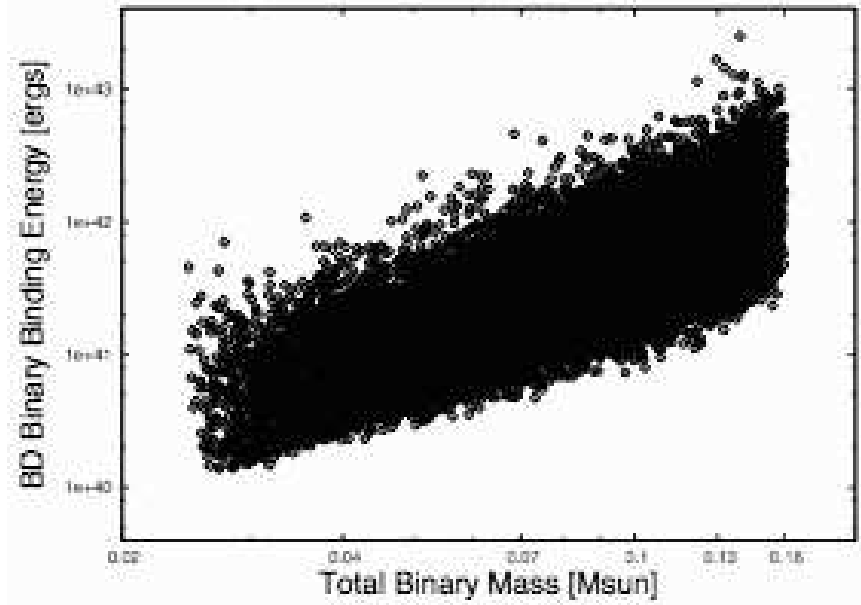}
\epsscale{1.0}
\caption{Binding energy for ejected brown dwarf binaries 
as function of total mass (M$_A$+M$_B$). 
The energies are measured in ergs.
\label{bindingenergy}}
\end{figure*}

\begin{figure*}
\epsscale{1.5}
\plotone{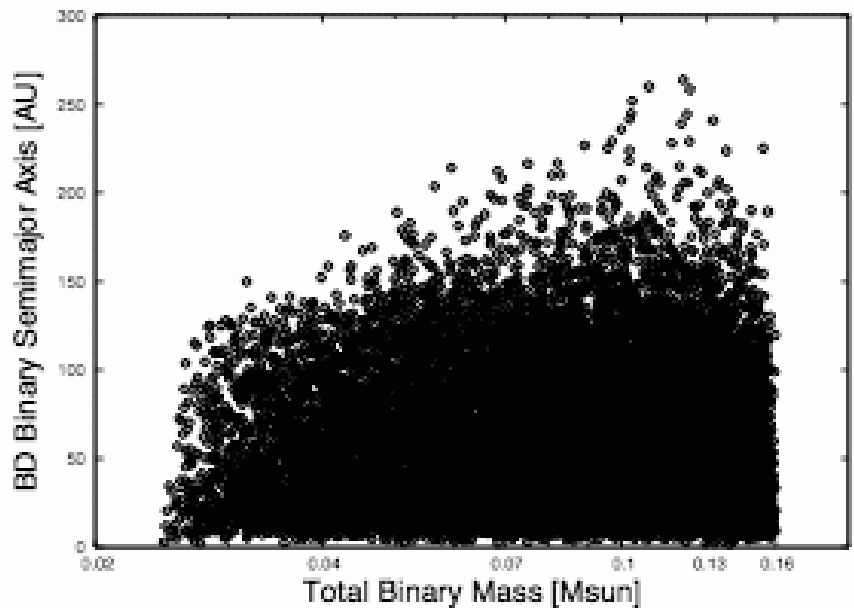}
\epsscale{1.0}
\caption{The semimajor axis of ejected BD binaries
as function of total binary mass.
\label{avsMab}}
\end{figure*}

Figure~\ref{bindingenergy} shows the binding energy as a function of
the total binary mass for all ejected BD binaries. The general trend
is that the binding energy increases with increasing mass.  This is
hardly surprising since the binding energy
($\propto$~$M_1$$M_2$/$a_{12}$) is directly proportional to the
individual component masses. The width of the distribution is partly
due to the fact that BD binaries have mass ratios from 1 down to 0.3
(see Figure~\ref{ejecteddiagnostic}) plus variations in the semimajor
axes. Figure~\ref{avsMab} shows the distribution of semimajor axes as
function of binary mass. As is well known from observations, and
reproduced by the simulations, mean binary semimajor axes increase
with binary mass. The effect is, however, much too small to counter
the growth of the binding energy due to increase in binary mass.  The
overall behavior of the binding energy fits well with the trends seen
in observations of VLM binaries (e.g., Close et al. 2007, Faherty et
al. 2011). As Figure~\ref{bindingenergy} shows, these BD binaries have
exceedingly small binding energies, as low as 10$^{40}$ ergs, which is
comparable to the values for the widest known binaries. While such
wide binaries are vulnerable to breakup by passing stars (e.g.,
Weinberg et al. 1987), the compact BD binaries are stable against
disruption, except possibly those that are born in dense clusters
(Close et al. 2007).

The binding energy is not a measure of the likelihood that a BD binary
could have survived an ejection event.  When a triple system breaks
up, a BD binary becomes permanently bound only at the moment when the
third body is ejected. The binding energy of the BD binary is set at
the moment of ejection and it can become high or low, depending on the
specific circumstances of ejection. It is thus not related to its {\em
  internal} survivability during system breakup. If there are more
bodies present, e.g. in a higher-order system or a cluster, then a
weakly bound binary is vulnerable to {\em external} perturbations, as
mentioned above.

\subsection{Kinematics of Binaries}
\label{sec:kinematics}

When a triple system breaks up and a single star escapes, then the
binary recoils in the opposite direction. It is of interest to ask if
the recoil leaves a measurable imprint on the binary space motion that
could be used to identify this violent event in its pre-history.
However, numerical studies of brown dwarfs ejected from disintegrating
triple systems show that while they may briefly during the actual
ejection process achieve a high velocity, they very soon slow down as
they climb out of the steep potential well of the cloud core and the
binary (Bate et al. 2003, Bate 2009). The mean
terminal velocity of all escapers in the 12,800 simulations by
Reipurth et al. (2010) is 1.13 km/s. Since the recoil velocity of the
binary is often lower because the binary tends to have a larger total
mass than the single, it follows that BD binaries formed through
disintegration have a velocity dispersion that is well within the
turbulent velocity range of star forming clouds, and they are thus
indistinguishable from BD binaries formed by other mechanisms.

\subsection{Survival of Circumstellar Material}
\label{sec:survival}

Reipurth \& Clarke (2001) noted that the dynamical interactions
involved in the ejection scenario implies that circumstellar material
around BDs will be truncated, and they suggested that ejected BDs
consequently should have less massive disks with weaker accretion
signatures and less or shorter-lived far-infrared excess emission.
Somehow, this has often been construed in the literature to mean that
young BDs that are ejected should not have disk-signatures like
infrared excesses. That is obviously not what was stated by Reipurth
\& Clarke (2001), nor what observations indicate (e.g., Monin et al.
2010, Bulger et al. 2014). While the outer regions of disks evidently
are truncated in a close triple approach (e.g., Hall et al.  1996),
this is a temporary situation since the viscosity of the disk material
ensures that the disks spread out again, allowing a full range of
temperatures, and accretion will also continue as long as disk
material remains.

Umbreit et al. (2011) have studied the effect of triple interactions
on the disks around BDs and the observable consequences. They find
that disks after close triple approaches are mostly less massive, but
compared to disks after two-body encounters they have similar or
larger radii. In a novel result, they also show that part of the disk
material initially lost is in fact re-captured by the ejected body
(see also, e.g., Steinhausen et al. 2012). Hydrodynamic simulations
like those of Bate (2003, 2009) are evidently important to study such
truncations in detail, although as pointed out by Bate (2012), one
should be aware that interactions in a large-N cluster may well be
different from the low-N systems we are considering here.

The same arguments that apply to single BDs are relevant for binary
BDs as well, in fact one may argue that the presence of a companion
provides another opportunity to capture otherwise escaping disk
material. It therefore seems that young BD binaries, just like young
single BDs, are likely to harbor circumstellar and/or circumbinary
disks, although loss of disk material might shorten the associated
signatures of youth.

\subsection{BD Binaries in the Field vs Clusters}
\label{sec:fieldclusters}

The very same processes described here for isolated triple systems are
almost certainly also taking place in clusters. The main difference in
the dynamical evolution of triple systems in loose associations and in
dense clusters is that the distant weakly bound components which in
low-density star forming regions in time will disrupt due to internal
instability, in clusters may be dislodged first due to
external perturbations. Parker \& Goodwin (2011) have shown that VLM
binaries with separations less than $\sim$20~AU are mostly immune to
disruption in even the densest clusters, while most VLM binaries with
separations of more than $\sim$100~AU can be destroyed in high-density
clusters, but are generally unaffected in low-density clusters. Since
most stars and BDs are formed in clusters, the effect is that the peak
and median values of the semimajor distribution of BD binaries
(Figure~\ref{semimajorclose}) will be lowered, in addition to the
effect expected from viscous interactions. This does not mean that
wide and very wide BD binaries necessarily must be rare, because
triple systems that eject a third body (i.e., dynamically ``unfold'')
{\em after} the initial clumpy substructure of clusters has been
smoothed out and the cluster starts to expand will have been largely
protected against disruption (Reipurth \& Mikkola 2012).

The main parameters that control the survival of a wide binary is thus
the unfolding time itself (which depends primarily on the semimajor
axis of the system), the cluster mass and relaxation timescale, and
the length of time since formation of the cluster until the triple
system begins transformation into a hierarchical configuration. Each
cluster will have its own signature on the binary/triple separation
distribution function, which will be truncated for separations
exceeding a cluster-dependent value (e.g., Kroupa et al. 1995,
Reipurth et al.  2007).

\section{BD BINARIES AS FUNCTION OF AGE:
SIMULATIONS VS OBSERVATIONS}
\label{sec:comparison}

\subsection{Properties of Eccentric Orbits}
\label{sec:properties}

\begin{figure*}
\epsscale{1.2}
\plotone{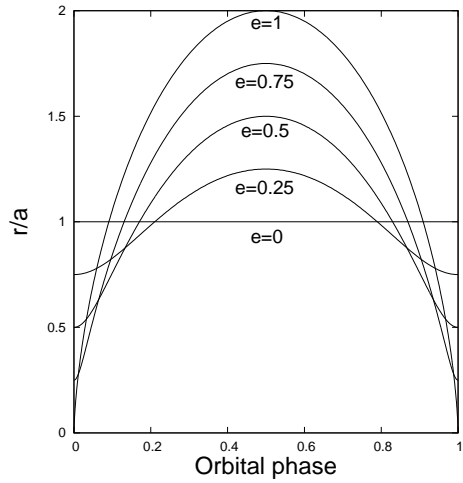}
\epsscale{1.0}
\caption{The separation of two binary components in units of the semimajor 
axis as a function of orbital phase for 5 different eccentricities 0 
(straight line), 0.25, 0.5, 0.75, and 1.
\label{separation}}
\end{figure*}

\begin{figure*}
\epsscale{1.2}
\plotone{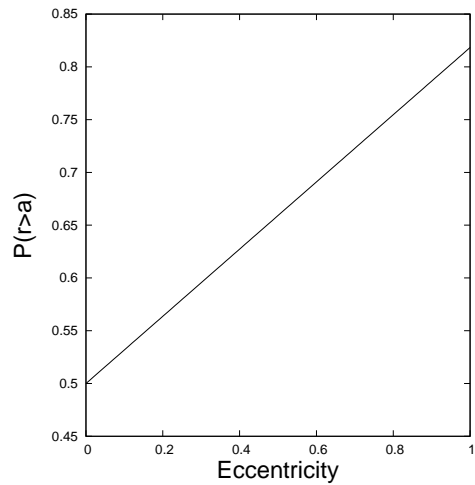}
\epsscale{1.0}
\caption{The fraction of time that a companion spends at separations $r$ 
larger than the semimajor axis $a$ as a function of eccentricity. For orbits 
that are nearly circular, the fraction is as expected
about 50\%, but for an eccentricity of 1 it is 82\%.
  \label{fractionoftime}}
\end{figure*}

Comparison of simulations with observations is hampered by the fact
that simulations frequently deal with semimajor axes {\em a} while
observations commonly only measure projected separations {\em s}.
Additionally, the orbital eccentricity plays a role. As demonstrated
by van Albada (1968), $<$log {\em s}$>$ -- $<$log {\em a}$>$ ranges
from $-$0.13 for circular orbits to as little as $-$0.03 for highly
eccentric orbits (see also Kuiper 1935, Couteau 1960, Halbwachs 1983).
When interpreting observations of individual binaries, it is sometimes
forgotten that for highly eccentric orbits the component separation
can be almost twice the semimajor axis, and in such systems the
fraction of time that a companion spends at separations larger than
the semimajor axis can reach 82\% (see Figures~\ref{separation} and
\ref{fractionoftime}). This is particularly important to consider when
dealing with systems originating from triple systems, which often
decay to highly eccentric orbits.

In the following we calculate the projected separations of the BD
binaries resulting from our simulations at different ages, allowing a
more direct comparison with observations.

\subsection{Projected Separations of Ejected BD Binaries as Function of Time}
\label{sec:BDprojsep}

Since we know the space coordinates of all our BD binaries at all
times, it is trivial to derive their projected separations.
Figure~\ref{BDejectedprojsep6} shows projected separations of the
9,209 BD binaries that have already been ejected at 1~Myr. Since none
of these binaries are suffering any perturbations following their
ejection, the distribution does not change with time, it is only the
number of binaries that increases as more unstable triples break up.
The distribution has a broad maximum between 2 and 20~AU, in which
35\% of all ejected BD binaries can be found. The peak is around
11~AU, and the median is at 28~AU. 

The projected distribution differs from the distribution of semimajor
axes (Figure~\ref{semimajorclose}) in two respects.  First, the random
distribution in space of the semimajor axes washes out the deep dip at
very small semimajor axes, which results because the simulations do
not include viscous interactions due to motion inside the gas core and
to disk-disk interactions at periastron.  The peak at 11~AU is
therefore an upper limit to the peak of BD binary separations.
Observationally, it is well established that binary separations
decline with diminishing stellar mass. Fisher \& Marcy (1992) found
that M-star binary separations peak around 4-30~AU, and Close et al.
(2003) and Maxted \& Jeffries (2005) both find a rather sharp peak of
VLM/BD objects around 4~AU, while Burgasser et al. (2007) estimated
the peak around $\sim$3-10~AU. This is, as expected, more compact than
our peak of projected separations around 11~AU due to the lack of
viscous effects in our simulations.  Second, the projected
distribution has a broader tail at larger separations, due to the fact
that for systems with high eccentricity, observers are likely to see a
system near apastron rather than periastron, as discussed above
(Section~\ref{sec:properties}).

\subsection{Dynamical State of BD Triple Systems in Star Forming
  Regions, in Moving Groups, and in the Field}
\label{sec:dynamicalstate}

\begin{figure*}
\epsscale{1.1}
\plotone{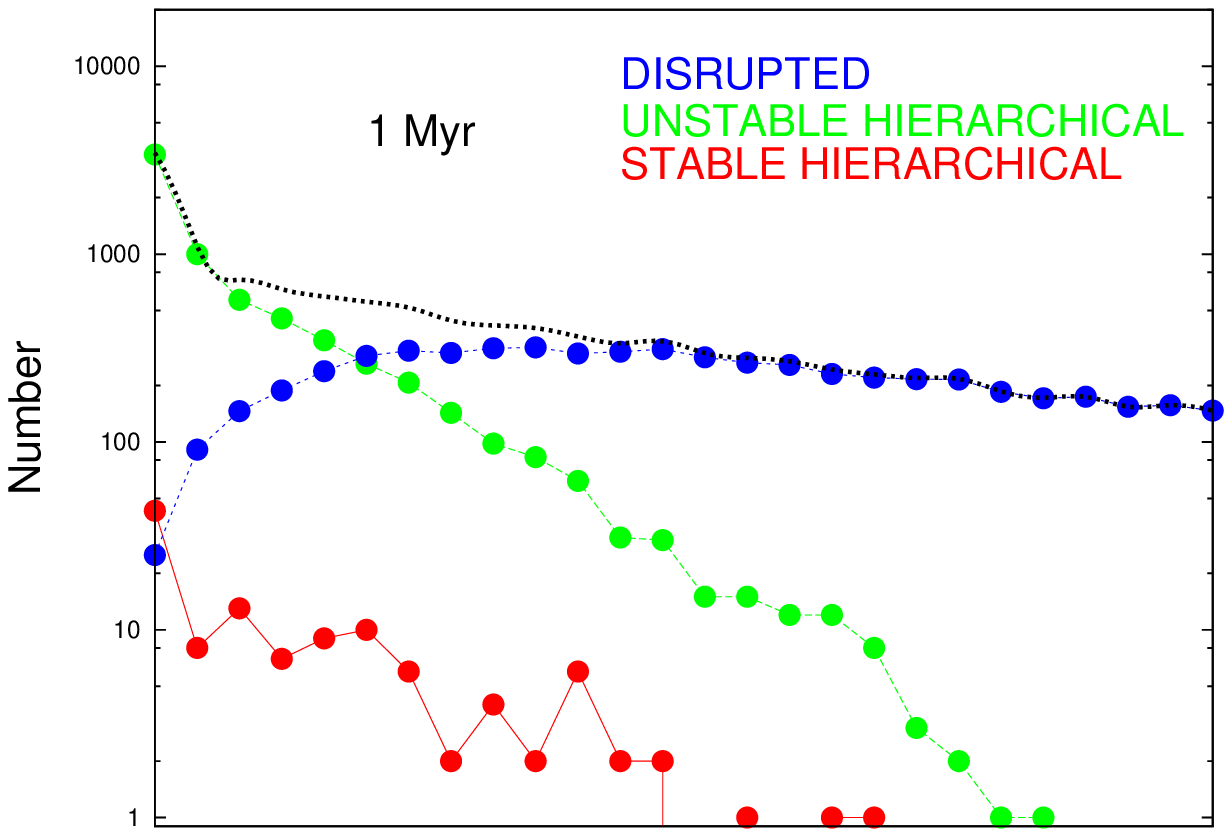}
\plotone{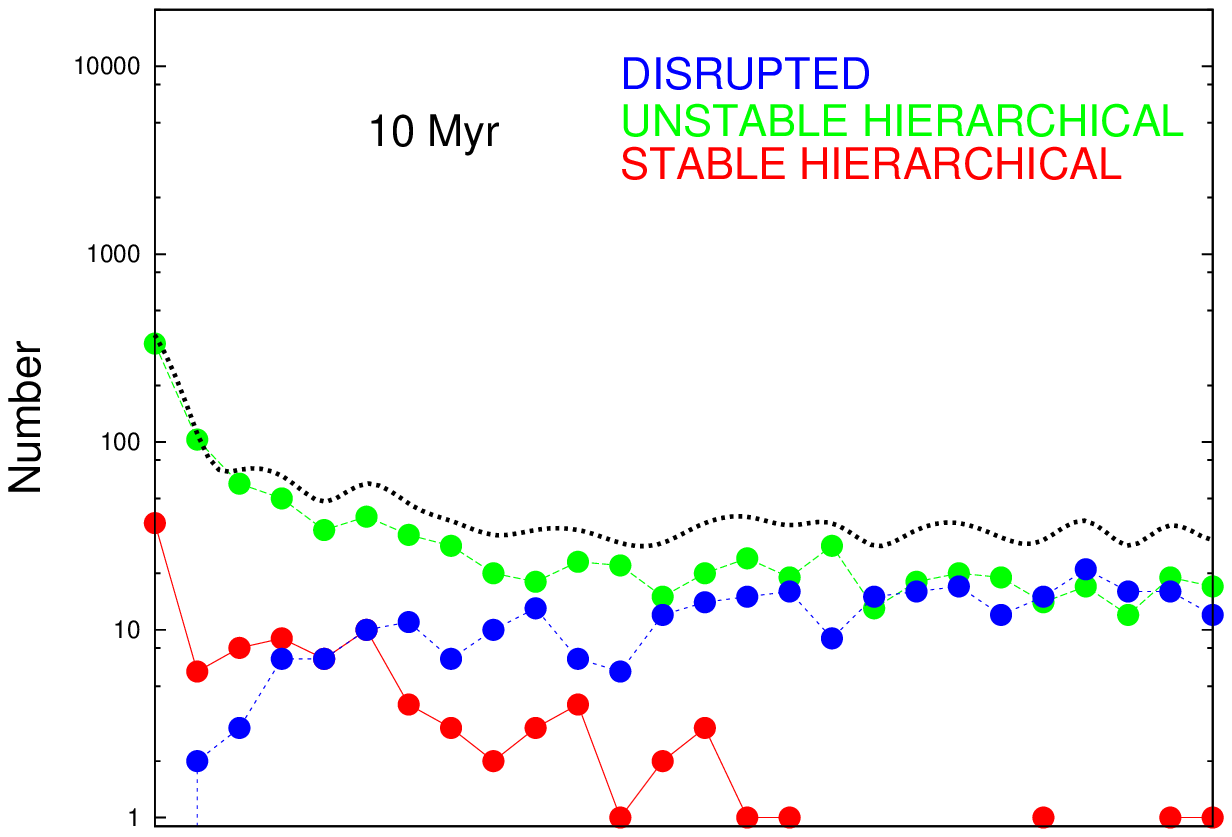}
\plotone{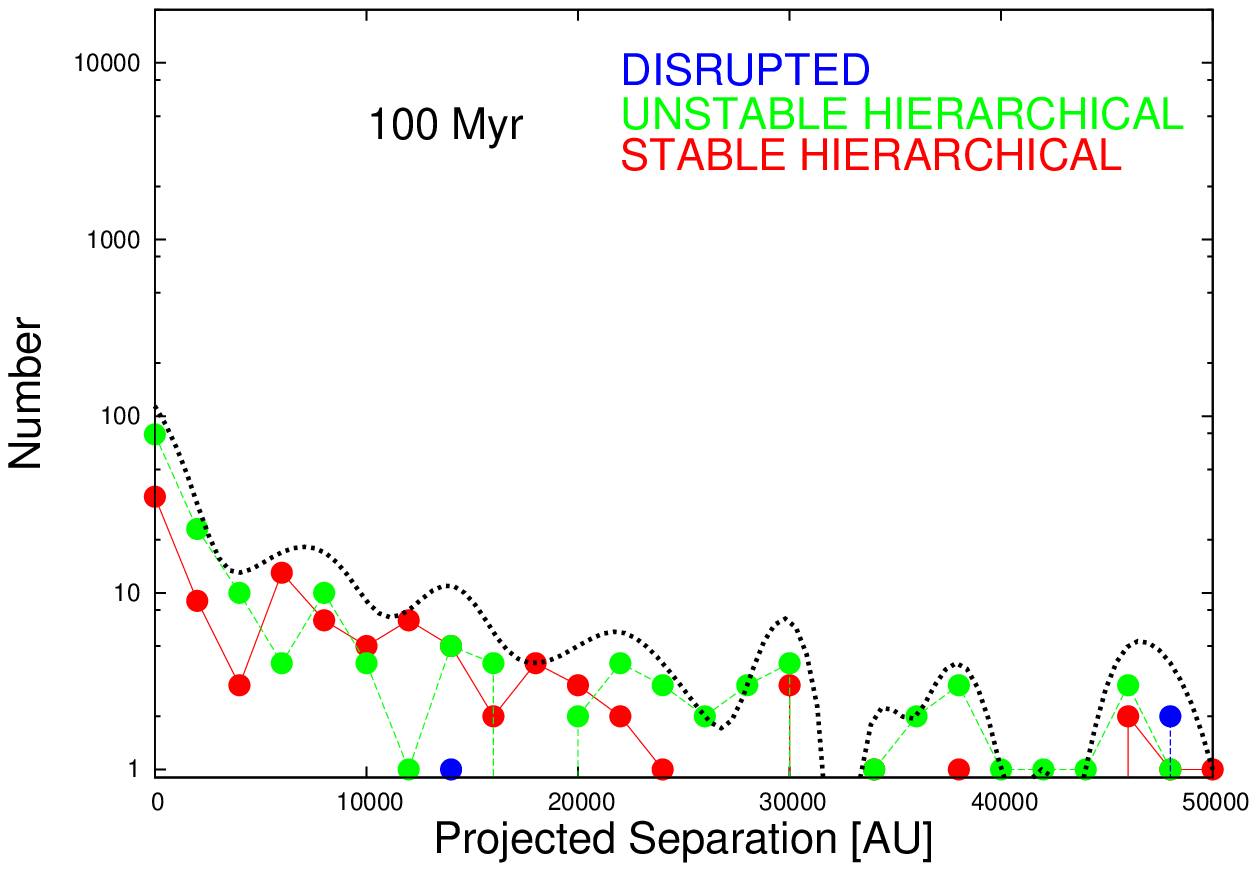}
\epsscale{1.0}
\caption{Total numbers of triple systems in which all three components are 
BDs are plotted as function of projected separation in intervals of 2,000~AU 
of the outer body relative to the center of mass of the inner binary. 
Bound triple systems are either stable (red) or unstable (green), and already 
disrupted systems are blue. The black dotted line indicates the sum of all
three categories as function of projected separation. A total of 15,376 BD
triple systems are plotted (including systems that fall outside the 
50,000~AU limit of the plot). The three panels show the projected 
separations at ages of 1, 10, and 100~Myr, respectively.
\label{projsep}}
\end{figure*}

The three panels in Figure~\ref{projsep} show the projected
separations of the outer body in triple systems relative to the center
of mass of the inner binary for systems in which all three components
are BDs at the ages of 1, 10, and 100~Myr, respectively. The systems
are divided into bound systems, which are either stable (red) or
unstable (green), or unbound systems in which the third body has
already been released from the system (blue).  The black dotted line
indicates the sum of all three types of systems.

\begin{figure*}
\epsscale{1.2}
\plotone{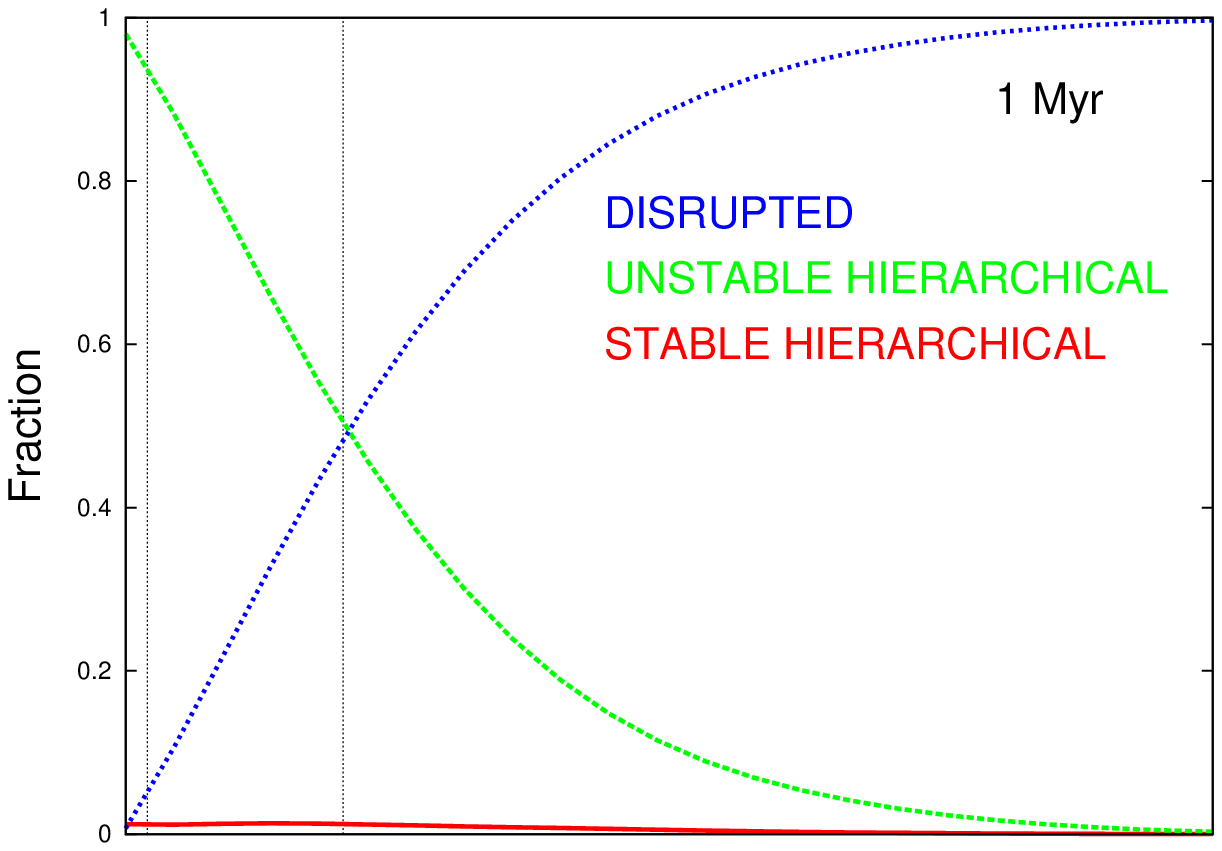}
\plotone{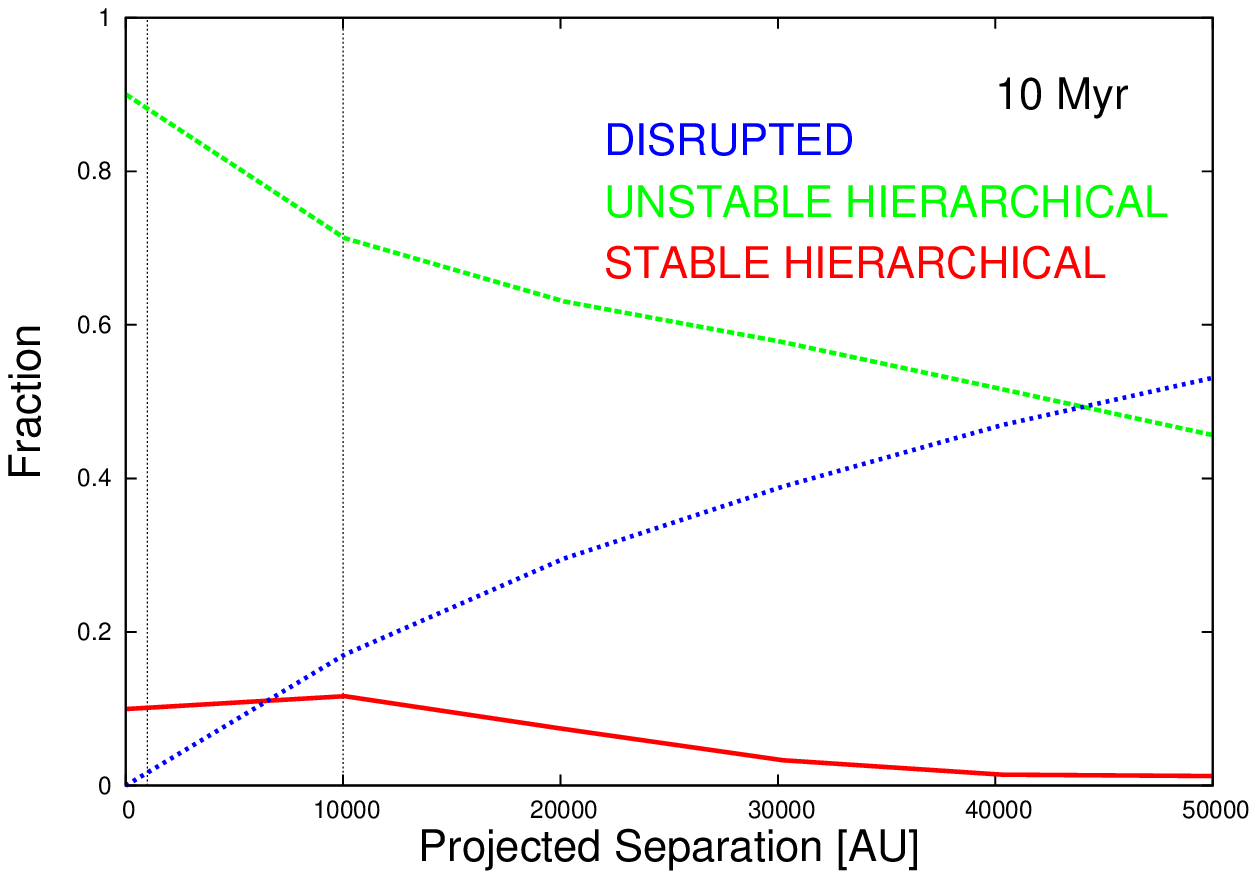}
\epsscale{1.0}
\caption{Fractions of stable, unstable, and disrupted BD triple systems 
are plotted as function of projected separation. The two vertical dotted lines 
mark separations of 1,000 and 10,000~AU, respectively. 
At 100~Myr the fractional values are dominated by small-number noise (see Figure~\ref{projsep} bottom) and are not shown.
\label{projsepfraction}}
\end{figure*}

At 1~Myr, BD triples with projected separations less than
approximately 10,000~AU are almost all still bound, but mostly
unstable.  At larger separations, the BD triple population is
dominated by disintegrated systems, where the third body is gently
traveling away from the binary to ever-increasing separations. Stable
triple systems consisting of three BDs are very rare, of the 15,376 BD
triples, only 117 systems are classified as having stable orbits at
1~Myr. The majority of these are compact, falling in the bin with
projected separations up to 2,000~AU, and only in this bin are the
stable bound systems more common than the already disintegrated
systems. The vast majority of BD triples are either bound but unstable
or already disrupted; out to about $\sim$10,000~AU the bound unstable
systems dominate the disrupted systems, but at larger separations this
reverses (see Figure~\ref{projsepfraction}).  At 1~Myr the number of
BD triple systems that are classified as stable, unstable, and
disrupted are 117, 6,748, and 8,511, corresponding to 0.7\%, 43.9\%,
and 55.4\%. Evidently, the breakup of triple systems is very common at
ages less than 1~Myr, as discussed by Reipurth et al.  (2010).

At 10~Myr, such as several populations in moving groups, significant
further dynamical evolution has taken place.  At this age, the number
of BD triple systems that are classified as stable, unstable, and
disrupted are 111, 1,365, and 13,900, corresponding to 0.7\%, 8.9\%, and
90.4\%. While almost all systems classified as stable at 1~Myr remains
so at 10~Myr, 80\% of the population of unstable triples at 1~Myr has
disrupted by 10~Myr. At the same time, the population of already
disrupted systems have moved even further apart, so in terms of
projected separations, most systems out to $\sim$45,000~AU are now
bound but unstable, and only at larger separations do the disrupted
systems become more common (see Figure~\ref{projsepfraction}). 

At 100~Myr, when almost all young groups have blended into the field
population, the number of BD triple systems that are classified as
stable, unstable, and disrupted are 110, 347, and 14,919, corresponding
to 0.7\%, 2.3\%, and 97.0\%. At even larger ages, it is likely that
the last unstable systems will have disintegrated. In other words, in
the field population only less than 1\% of BD triple systems should
still be bound.

It should be kept in mind that in many BD triple systems, viscous
interactions in the close binary will cause a spiral-in of the
components so they form a spectroscopic binary, and in some cases will
even merge (see Section~\ref{sec:spectroscopicbinaries}). Such systems
will appear as wide BD binaries. It is therefore of interest to note
that Close et al. (2007) have estimated that $\sim$6\%$\pm$3\% of
young VLM objects are found in wide ($>$100~AU) systems, while only
0.3\%$\pm$0.1\% of old field VLM objects are found in such wide
systems. This finding is supported by the more recent observations of
Todorov et al. (2014).

\subsection{BD Binaries associated with Main-sequence Stars}
\label{sec:mainsequencestars}

\begin{figure*}
\epsscale{1.0}
\plotone{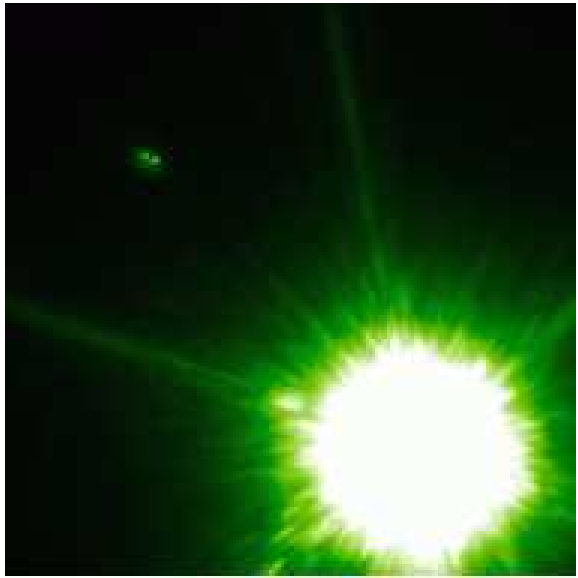}
\epsscale{1.0}
\caption{A brown dwarf binary (L4+L4) is associated with the G2~V star
  HD~130948 (Potter et al. 2002).  The system has an age of about 800 million 
  years, and at
  a distance of 18~pc the projected separation between primary and the binary 
  is $\sim$47~AU (Dupuy, Liu, \& Ireland 2009). Such systems can form from 
  three identical stellar embryos, where one by chance gains more mass and 
  dynamically banish the other two to the outskirts of the cloud core, where
  they never gain enough mass to start hydrogen burning.
  Courtesy Trent Dupuy \& Michael Liu.
  \label{hd130948}}
\end{figure*}

It is well known that BD binaries in some cases are found as
companions to main sequence or evolved stars. A fine example is the
G2V star HD~130948, which has a L4+L4 pair at a projected separation
of 47~AU from the star and orbiting the star with a very high
eccentricity, with the eccentricity probability function peaking at
0.83 (Potter et al.  2002, Dupuy et al.  2009, Ginski et al. 2013),
see Figure~\ref{hd130948}. The BD binary itself has a projected
separation of $\sim$2.4~AU and an orbital period of about 10~yr (Dupuy
\& Liu 2011).

We here argue that such BD binary companions are a straightforward
consequence of dynamical interactions in triple systems. Basically,
such stars with BD binary companions are the triple systems where the
BD binary was not successfully ejected into an escape, but was ejected
into a stable orbit around the main component. Our simulations show
that such systems can originate from three identical stellar embryos,
where one by chance has controlled a position near the center of the
molecular core and rapidly grew to stellar mass while dynamically
keeping the other two components in the outskirts of the core, where
they were unable to gain much mass.

Such systems are rare, and become even rarer as the systems evolve,
since the majority are unstable and therefore much dynamical evolution
takes place at early stages. At 1~Myr, there are 1,325 BD binaries
bound to a star among our 200,000 simulations, corresponding to
0.66\%. By 10~Myr less than half remain (582, corresponding to
0.29\%).  At 100~Myr only 377 are left (0.19\%), and of these 253
(0.13\%) are stable and are likely to remain so on much longer
timescales.

The number of BD binaries bound to a star decreases with time, while
the number of ejected BD binaries increases with time. Hence the ratio
of BD binaries bound to a star relative to the number of free-floating
already ejected BD binaries varies with time, at 1~Myr it is
1,325/9,209 ($\sim$1:7), at 10~Myr 582/14,814 ($\sim$1:25), and at 100~Myr
377/15,894 ($\sim$1:42).

The large majority of BD binaries bound in a triple system are bound
to another BD, not to a star. So if we ask the more general
question of what fraction of all BD binaries that are bound (to either
a star or a BD) relative to all BD binaries (bound plus unbound) we
get 47.1\% at 1~Myr, 12.2\% at 10~Myr, and 5.0\% at 100~Myr.

\begin{figure*}
\epsscale{1.5}
\plotone{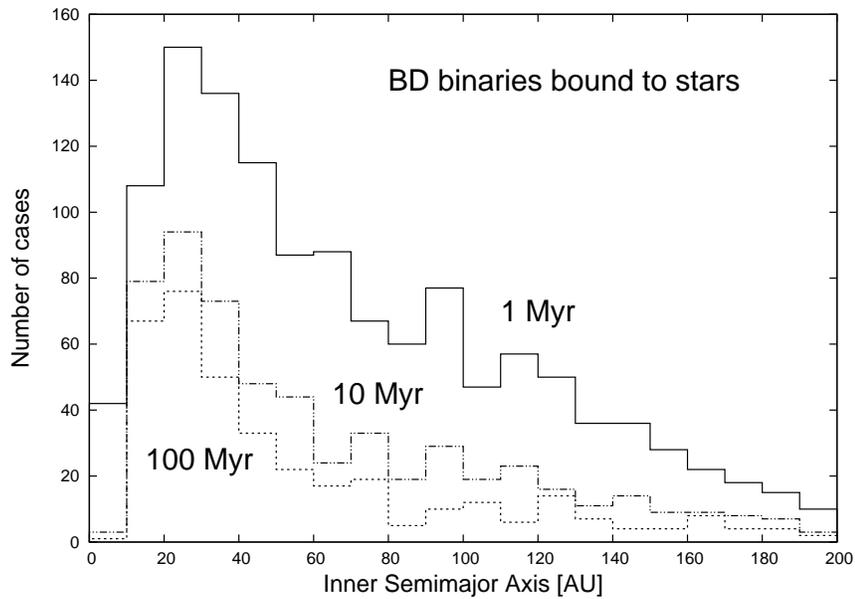}
\epsscale{1.0}
\caption{Semi-major axis distribution of brown dwarf binaries bound to stars 
at 1~Myr, 10~Myr, and 100~Myr. At those ages there are 1,325, 582, and 377 
BD binaries bound to stars.
\label{semimajorclosebound}}
%
%
%

\end{figure*}

\begin{figure*}
\epsscale{1.5}
\plotone{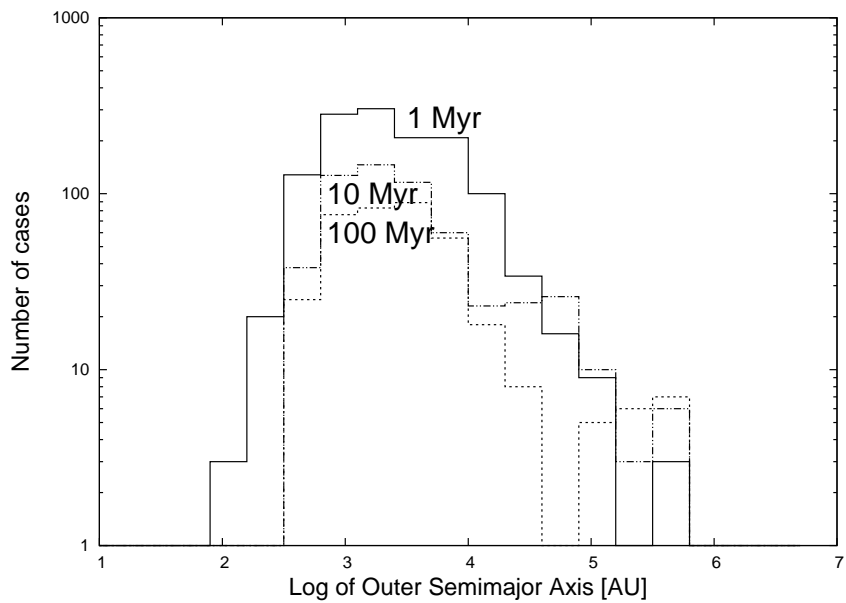}
\epsscale{1.0}
\caption{Semi-major axis distribution of the outer components in triple systems
consisting of a BD binary bound to a star, at 1~Myr, 10~Myr, and 100~Myr. 
These are the same systems as in Figure~\ref{semimajorclosebound}. 
The figure shows that the majority of BD binaries that remain bound to stars 
are having semimajor axes of a thousand AU or larger.
\label{semimajordistantbound}}
%
%
%
\end{figure*}

The semimajor axis distribution of the BD binaries bound to stars is
seen in Figure~\ref{semimajorclosebound}. The peak of the distribution
is around 25~AU, and the median value is about 59~AU. The large
majority of these bound BD binaries are orbiting the central star on
wide orbits, with semimajor axes of typically thousands of AU
(Figure~\ref{semimajordistantbound}). 

Only a small number of BD binaries bound to stars have been discovered
so far, and with such small-number-statistics, it is difficult to make
a meaningful comparison between observations and simulations. But the
existence of a BD binary associated with $\epsilon$~Indi (e.g., King
et al.  2010), which is located at a distance of 3.6~pc and is one of
only about 25 stars this close, may suggest that such BD binary
companions are not extremely rare.  Comparison with observations is
complicated by the fact that many of these bound BD binaries require
adaptive optics imaging to be resolved, which in many cases limit
discoveries to binaries with projected separations to their host star
of only a few arcseconds. As widefield adaptive optics systems become
more common, more wide systems are expected to be discovered, enabling
meaningful comparisons between simulations and observations. From the
limited observations available, Burgasser et al. (2005) and Faherty et
al. (2010, 2011) suggest that stars are more than twice as likely to
have a brown dwarf binary rather than a single BD as a companion,
which strongly points to a dynamical origin of BD binary companions to
stars.

\section{LIMITATIONS OF  PRESENT STUDY}
\label{sec:limitations}

In this paper we have studied the motion of triple systems initially
inside the dense cloud cores from which they formed, and we have
explored the important interplay between accretion and orbital
dynamics, which alters the properties of the resulting single, binary,
and triple systems. Our study is only a first step, and future studies
can improve on these results in several ways, some of which we
identify below.

$\bullet$ In order to achieve the 200,000 simulations required to
obtain solid statistics of the outcomes, we have been forced by
limitations in computer power to use Bondi-Hoyle accretion. Clearly,
the use of smoothed particle hydrodynamics would be a major step
forward (e.g., Delgado-Donate et al. 2004, Hubber et al. 2013).  Also,
eventually inclusion of magnetic fields will be necessary, since
magnetic braking is efficient in removing angular momentum of the
infalling gas, and thus tightens a resulting binary or higher-order
multiple (Zhao \& Li 2013).

$\bullet$ While we include the braking effect of the gas on the motion
of the triple components, we treat the bodies as point sources, and
specifically we do not include circumstellar material. The presence of
dense gas in circumstellar disks and envelopes will cause viscous
interactions, particularly during periastron passages, which will
initiate inspiral phases (e.g., Korntreff et al. 2012). This is the
mechanism by which close spectroscopic binaries are formed. In the
more extreme cases, mergers are likely to take place.

$\bullet$ Our use of an initial mass function is an approximation in
two ways. First, we employ an IMF that is truncated between 0.012 and
2 M$_\odot$. Second, we start with an observed IMF and then allow the
stars to accrete, which alters the IMF. Ideally, one would guess a
mass distribution of the stellar embryos that would end up with the
observed IMF. Given our current complete ignorance about the
properties of stellar seeds, we have made no such attempts.

$\bullet$ A number of assumptions for the simulations were made which
have limited or no constraints from observations. We set the initial
separations between the bodies to be from 40 to 400~AU; these numbers
will be better constrained as centimeter and millimeter
interferometric observations of protostars become more common. We have
assumed that the three embryos are always identical in mass and are
born simultaneously, but both of these assumptions may turn out to not
always be true. Finally, we have no knowledge of whether the gas has
any angular momentum, and have assumed none. However, if the gas has
angular momentum, this may alter the separation of the systems (e.g.,
Bate 2000, Umbreit et al. 2005).

$\bullet$ In this paper we have focused on the dynamical interactions
in triple systems. But observations show that stable higher-order
systems are relatively common (e.g., Raghavan et al. 2010). Given that
the stability of higher-order systems is considerably more difficult
to achieve than for triple systems, we infer that many more
higher-order systems are born than are observed in the field.
Exploratory simulations of higher-order systems in gas clouds show
that they frequently decay by ejecting their lowest mass members
including many brown dwarfs. This will lower the unrealistically high
binary fraction among BDs we find in this paper and bring it more in
accordance with observations.  Detailed simulations of higher-order
systems are required to quantify this effect.

In short, there is room for significant improvements to what we have
presented here. We see the goal of the present study to be laying the
groundwork for future more detailed studies of the importance of
dynamical evolution of multiple systems coupled with accretion, which
we believe is of central importance for understanding many properties
of stellar populations.

\section{CONCLUSIONS}
\label{sec:conclusions}

We have performed 200,000 N-body simulations with three identical
stellar seeds drawn from an IMF and embedded in a cloud core, allowing
accretion onto the bodies. The following main results were obtained:

{\em 1.} Non-hierarchical triple systems are strongly affected by
their environment, and accretion from the surrounding gas alters the
individual masses. Random motions in the centrally condensed gas cloud
introduce differences in mass that allow one or two of the stellar
embryos to control the gas rich center and dynamically banish the
other triple member(s) to the outskirts of the gas cloud, where
it/they can accrete only very little. The combination of dynamics and
accretion thus introduces important differences among the members of a
triple system.

{\em 2.} The large majority of triple systems break up, and the
simulations show that not only single substellar objects are ejected
into escapes, but that free-floating BD binaries are a common and
natural consequence of the disintegration of triple systems moving and
accreting inside a gas cloud.

{\em 3.} In order to characterize the population of triple systems
(bound and disrupted) that result from the simulations, we have
designed the 'triple diagnostic diagram', which plots two
dimensionless numbers against each other, representing the mass ratio
of the binary vs the mass ratio of the third body relative to the
total system mass.

{\em 4.} The separation distribution function of ejected BD binaries
bears a strong resemblance to the current limited observations of BD
binaries, with a broad peak of the projected separations between 2 and
20~AU. The ejected BD binaries tend to have high eccentricities, but
the simulations do not include viscous interactions, which may cause
spiral-in and a decrease of the eccentricity. Current observations
suggest a BD binary fraction of 20\% - 45\%. The simulations appear to
overproduce BD binaries (fraction 0.43), but this ignores the breakup
of higher-order multiples, which produce a higher number of single
BDs, thus lowering the overall BD binary fraction.

{\em 5.} Dynamical evolution in triple systems does not produce BD
binaries with separations exceeding $\sim$250~AU. Wide and very wide
BD binaries formed through ejection must therefore represent bound BD
triple systems where the close pair is either unresolved or has merged
as a result of viscous interactions in the protostellar phase.

{\em 6.} Dynamical evolution is rapid and in many cases brief, and
multiplicity as measured in even the youngest star forming regions is
thus not representative of the primordial multiplicity, since more
than half of all systems disintegrate already during the protostellar
stage.

{\em 7.} The large majority of bound triple systems are unstable,
eventually leading to breakup, so the main threat to bound triple
systems is not from external perturbations but is due to internal
instabilities.

\acknowledgments

We are grateful to an anonymous referee, to Matthew Bate, and to
Hsin-Fang Chiang for very helpful comments on the manuscript.  BR
thanks Andrei Tokovinin and Mark Chun for advice on IDL routines,
Hsin-Fang Chiang and Colin Aspin for providing additional computer
power to complete these simulations, and Tuorla Observatory for
hospitality during several visits.  This material is based upon work
supported by the National Aeronautics and Space Administration through
the NASA Astrobiology Institute under Cooperative Agreement No.
NNA09DA77A issued through the Office of Space Science.  This research
has made use of the SIMBAD database, operated at CDS, Strasbourg,
France, and of NASA's Astrophysics Data System Bibliographic Services.


\clearpage


\clearpage

\vspace{0.3cm}

\clearpage

\clearpage

\clearpage








\clearpage

\clearpage
    
\clearpage

\clearpage

\clearpage
    
\clearpage
    
\clearpage
    
\vspace{0.5cm}
    
\clearpage


\clearpage

\clearpage

%
%
%
%

\clearpage

\clearpage
    
\clearpage

\begin{deluxetable}{ccccrrrrrrrrrrrr}
\tabletypesize{\normalsize}
\tablecaption{Percentages of different triple systems with stars and brown dwarfs.$^a$ 
\label{tbl-2}}
\tablewidth{0pt}
\tablehead{
\multicolumn{3}{c}{Components$^b$} & \colhead{$|$} & \multicolumn{3}{c}{1 Myr} & & \multicolumn{3}{c}{10 Myr} & & \multicolumn{3}{c}{100 Myr}\\
\cline{1-3} \cline{5-7} \cline{9-11} \cline{13-15}
\colhead{A} & \colhead{B} & \colhead{C} & \colhead{$|$}  & \colhead{H} & \colhead{U} & \colhead{D} & & \colhead{H} & \colhead{U} & \colhead{D} & & \colhead{H} & \colhead{U} & \colhead{D} 
}
\startdata
BD & BD & BD          & $|$ & 0.06 & 3.37 & 4.26 && 0.06 & 0.68 & 6.95 && 0.06 & 0.17 & 7.46 \\
$\star$ & BD & BD      & $|$ & 0.06 & 0.93 & 1.41 && 0.06 & 0.27 & 2.33 && 0.06 & 0.07 & 2.61 \\
$\star$ & $\star$ & BD  & $|$ & 0.01 & 0.24 & 0.40 && 0.01 & 0.05 & 0.67 && 0.01 & 0.01 & 0.72 \\
BD & BD & $\star$       & $|$ & 0.13 & 0.53 & 0.35 && 0.13 & 0.16 & 0.45 && 0.13 & 0.06 & 0.48 \\
$\star$ & BD & $\star$   & $|$ & 0.07 & 0.37 & 0.35 && 0.07 & 0.12 & 0.52 && 0.07 & 0.06 & 0.57 \\
$\star$ &$\star$ &$\star$ & $|$ & 7.43 & 25.84 & 54.14 && 7.29 & 9.32 & 70.80 && 6.95 & 4.28 & 76.18 \\
\cline{1-15}
 & All &  & $|$  & 7.76 & 31.28 & 60.91 && 7.62 & 10.60 & 81.72 && 7.28 & 4.65 & 88.02 \\
\enddata
\tablenotetext{a}{H: stable hierarchical - U: unstable hierarchical - D: disrupted}
\tablenotetext{b}{A and B form a binary, C is the third body either distantly bound or escaped} 


\end{deluxetable}

\begin{deluxetable}{lccccc}
\tabletypesize{\normalsize}
\tablecaption{Single vs binary ejected brown dwarfs at 1 Myr$^a$  
\label{tbl-2}}
\tablewidth{0pt}
\tablehead{
\colhead{} & \colhead{$|$} & \multicolumn{1}{c}{All binaries} &  \multicolumn{1}{c}{25\% binaries} &  \multicolumn{1}{c}{50\% binaries} & \multicolumn{1}{c}{75\% binaries}\\
\colhead{} & \colhead{$|$}  & \multicolumn{1}{c}{fully resolved$^b$} & \multicolumn{1}{c}{not resolved$^c$} & \multicolumn{1}{c}{not resolved$^c$} & \multicolumn{1}{c}{not resolved$^c$} 
}
\startdata
BD Singles:             & $|$ &  6.07\%  &  7.22\%  &  8.37\%  &  9.53\% \\
BD Binaries:            & $|$ &  4.61\%  &  3.46\%  &  2.31\%  &  1.15\% \\
BD Total:               & $|$ & 10.68\%  & 10.68\%  & 10.68\%  & 10.68\% \\
BD Binary fraction:$^d$ & $|$ &  0.43    &  0.32    &  0.22    &  0.11   \\
\enddata

\tablenotetext{a}{Percentages of simulations that produce ejected
  single or binary brown dwarfs} 
\tablenotetext{b}{This assumes observations can resolve {\em all} binaries}
\tablenotetext{c}{This assumes observations fail to resolve some binaries, and count them as singles}
\tablenotetext{d}{The numbers refer to simulations at 1 Myr. The total number
of (single + binary) brown dwarfs ejected at 10 Myr is 17.35\% and at 100 Myr is 18.73\% 
The (fully resolved) binary fraction remains constant at 0.43 at 1, 10, and 100~Myr.}
\end{deluxetable}

\end{document}